\documentclass[lettersize,journal]{IEEEtran}

\usepackage{amsmath,amssymb,amsfonts}
\usepackage{algorithmic}
\usepackage{graphicx}
\usepackage{float}
\usepackage{subfigure}
\usepackage{bm}
\usepackage{textcomp}
\usepackage{color}
\usepackage[linesnumbered, ruled]{algorithm2e}
\usepackage{verbatim}
\usepackage{amsmath,lipsum}
\usepackage{cuted}
\usepackage{xcolor}
 
\ifCLASSOPTIONcompsoc
\usepackage[caption=false,font=normalsize,labelfon
t=sf,textfont=sf]{subfig}
\else
\usepackage[caption=false,font=footnotesize]{subfig}
\fi

\ifCLASSOPTIONcompsoc
  \usepackage[nocompress]{cite}
\else
  \usepackage{cite}
\fi

\ifCLASSINFOpdf
\else
\fi

\usepackage{multirow}
\usepackage{bm}
\usepackage{booktabs}
\usepackage{ntheorem}
\usepackage{tabularx}
\theoremseparator{.}
\newskip\theorempreskipamount
\newskip\theorempostskipamount

\usepackage{tabularx}
\usepackage{amssymb}
\usepackage{pifont}
\theorembodyfont{}

\allowdisplaybreaks[4]

\begin{document}

\title{Dual UAV Cluster-Assisted Maritime Physical Layer Secure Communications via Collaborative Beamforming}

\author{Jiawei Huang,
        Aimin Wang,
        Geng Sun,~\IEEEmembership{Senior Member,~IEEE,}
        Jiahui Li,
        Jiacheng Wang,
        Hongyang Du,\\
        and Dusit Niyato,~\IEEEmembership{Fellow,~IEEE}
\thanks{Jiawei Huang, Aimin Wang and Jiahui Li are with the College of Computer Science and Technology, Jilin University, Changchun 130012, China, and Key Laboratory of Symbolic Computation and Knowledge Engineering of Ministry of Education, Jilin University, Changchun 130012, China (E-mail: huangjiawei97@foxmail.com, wangam@jlu.edu.cn, lijiahui@jlu.edu.cn).}
\thanks{Geng Sun is with the College of Computer Science and Technology, Jilin University, Changchun 130012, China, and also with the Key Laboratory of Symbolic Computation and Knowledge Engineering of Ministry of Education, Jilin University, Changchun 130012, China. He is also with the College of Computing and Data Science, Nanyang Technological University, Singapore 639798 (E-mail: sungeng@jlu.edu.cn).}
\thanks{Jiacheng Wang and Dusit Niyato are with the College of Computing and Data Science, Nanyang Technological University, Singapore 639798 (E-mail: jiacheng.wang@ntu.edu.sg, dniyato@ntu.edu.sg).}
\thanks{Hongyang Du is with the Department of Electrical and Electronic Engineering, Hong Kong University, Hong Kong 999077 (E-mail: duhy@eee.hku.hk).}

\thanks{Part of this paper appeared in IEEE CSCWD 2023~\cite{Huang2023}.}
\thanks{\textit{(Corresponding author: Geng Sun and Jiahui Li.)}}
}

\markboth{Journal of \LaTeX\ Class Files,~Vol.~14, No.~8, August~2021}%
{Shell \MakeLowercase{\textit{et al.}}: A Sample Article Using IEEEtran.cls for IEEE Journals}

\maketitle

\begin{abstract}
Unmanned aerial vehicles (UAVs) can be utilized as relay platforms to assist maritime wireless communications. However, complex channels and multipath effects at sea can adversely affect the quality of UAV transmitted signals. Collaborative beamforming (CB) can enhance the signal strength and range to assist the UAV relay for remote maritime communications. However, due to the open nature of UAV channels, security issue requires special consideration. This paper proposes a dual UAV cluster-assisted system via CB to achieve physical layer security in maritime wireless communications. Specifically, one UAV cluster forms a maritime UAV-enabled virtual antenna array (MUVAA) relay to forward data signals to the remote legitimate vessel, and the other UAV cluster forms an MUVAA jammer to send jamming signals to the remote eavesdropper. In this system, we formulate a secure and energy-efficient maritime communication multi-objective optimization problem (SEMCMOP) to maximize the signal-to-interference-plus-noise ratio (SINR) of the legitimate vessel, minimize the SINR of the eavesdropping vessel and minimize the total flight energy consumption of UAVs. Since the SEMCMOP is an NP-hard and large-scale optimization problem, we propose an improved swarm intelligence optimization algorithm with chaotic solution initialization and hybrid solution update strategies to solve the problem. Simulation results indicate that the proposed algorithm outperforms other comparison algorithms, and it can achieve more efficient signal transmission by using the CB-based method.
\end{abstract}


%

\section{Introduction}
\label{sec:introduction}

\IEEEPARstart {I}{n} recent years, owing to the continuous development of the marine economy, marine services have a wide range of applications in military, civilian, and commercial fields, and it is urgent to establish an efficient and reliable maritime communication network~\cite{Zhang2024}~\cite{Fang2023}. However, the challenge of installing communications equipment at sea results in lower maritime signal transmission rates than cellular networks at present~\cite{Wang2022}. Due to the advantages of wide coverage, simple deployment, and low cost, unmanned aerial vehicles (UAVs) can be regarded as effective relay platforms to assist maritime wireless communications~\cite{Nomikos2024}~\cite{Qian2023}. However, the higher flight altitude of UAV brings long-distance communications, which may render gradual signal attenuation during the propagation, further impacting the effectiveness and reliability of the communication link.

\par Collaborative beamforming (CB) is considered to be a promising method that can enhance the transmission performance of UAVs as a relay.  Specifically, multiple array elements on a UAV cluster can form a virtual antenna array (VAA). Then, VAA transmits synchronously among the array elements so that constructive signals are available at the location of the receiving user~\cite{Sun2022}. Ideally, $N_{U}$ array elements in the VAA can generate $N_{U}^{2}$ times gain to the target via CB~\cite{Jayaprakasam2017}. Therefore, CB can improve the communication performance of UAVs at high altitudes without changing the equipment. However, the open channel of the UAVs, as well as the increased transmission range, make the signals more susceptible to malicious eavesdropping~\cite{Zhu2020},~\cite{Bastami2021}.

\par The conventional upper layer decryption and encryption methods require high computing ability for frequent encoding and decoding, which is extremely challenging for resource-limited UAVs and marine services~\cite{Mahmood2023}. Different from these methods, physical layer security (PLS) is an effective way to accomplish secure wireless communication due to its strong adaptability~\cite{Wu2018}~\cite{Wang2019}. Moreover, the flexibility of UAVs has made them increasingly attractive for maritime PLS applications~\cite{Zhou2018}. For example, Dang \textit{et al.}~\cite{dang2022secure} presented a UAV-aided friendly-jamming architecture to strengthen safety performance by adjusting the positions of the UAVs. Liu \textit{et al.}~\cite{Liu2021} designed a maritime anti-jamming transmission framework by using UAVs, and optimized the moving path and power allocation of the UAV to improve the performance. However, the abovementioned power allocation method can decrease the communication rates of legitimate users. Furthermore, these works need the UAVs to fly a long distance from original locations to target locations~\cite{Wu2021a}, which consumes much energy. Note that the energy consumption needs to be focused, since it is a critical factor in realistic maritime communications and determines the communication duration. In addition, due to the complexity of the maritime channels, UAVs are more likely to crash when they fly far away from the shore or vessel.

\par Similarly, based on the long-range transmission characteristic of CB, another set of UAVs can form a VAA as jammer to send jamming signals directly to the remote eavesdropping vessel, protecting data from decoding through noise jamming. In this case, UAVs can form two communication types for long-range and friendly-jamming secure maritime wireless communications. Specifically, one UAV cluster forms a maritime UAV-enabled virtual antenna array (MUVAA) relay to forward data signals to the legitimate vessel by CB, and the other UAV cluster forms an MUVAA jammer to send jamming signals to the eavesdropper via CB. The CB-based system can be applied to realistic scenarios. For example, vessels are difficult to approach in a disaster situation at sea as the existence of obstacles and limitation of routes~\cite{Rani2023}. In this case, the VAA can use CB to send real-time data signals to the rescue vessel, and another distant VAA can protect against eavesdropping by sending jamming signals through the CB.

\par Note that the jamming signals may also be sent to the area of legitimate users, reducing the communication quality of data signals. Thus, we need to precisely design the VAA to reduce the undesirable impact of jamming signals. Since the performance of the VAA is determined by the 3D positions and excitation current weights of UAVs, making CB implementation complex in such systems. Hence, we require jointly control the positions and excitation current weights of UAVs in the MUVAA relay and MUVAA jammer to optimize the performance of maritime communications. However, this process involves a large number of variables. Moreover, the process of position adjustment also incurs energy consumption, which means that maritime communication efficiency and UAV energy usage are conflicted. Consequently, achieving the trade-off between the transmission efficiency of VAA and the total flight energy consumption of UAVs is a challenge. In this case, swarm intelligence algorithms have recently advanced, improving their global search capabilities and convergence rates~\cite{Tang2023}. In the multi-objective optimization problem (MOP), these algorithms excel by efficiently exploring solution spaces to identify Pareto optimal sets and balancing conflicting multiple objectives. Their parallel search and diversity maintenance make them effective for non-linear MOP~\cite{Kuo2023}.

\par As far as we know, this is the first work to consider the dual UAV cluster-assisted maritime secure communications via CB, analyze conflicts between multiple objectives, and present a novel swarm intelligence algorithm to resolve them. The primary contributions of this paper are summarized as follows.

\begin{itemize}
\item \textbf{\textit{CB-based Dual UAV Cluster-Assisted Maritime Secure Communication System:}} We propose using one UAV cluster to form an MUVAA relay, which can forward data signals to the remote legitimate vessel directly via CB. Then, the other UAV cluster forms an MUVAA jammer which can send jamming signals directly to the remote illegitimate user by CB to protect against eavesdropping. The system can facilitate maritime wireless communications and ensure security while reducing the flight distance of the UAVs, thus improving energy efficiency.

\item \textbf{\textit{Multi-objective Optimization Problem Formulation:}} Considering that the implementation of maritime secure communications and the energy consumption of a UAV are in conflict with each other, we adopt a multi-objective optimization scheme to trade off the optimization objectives. Thus, we formulate a secure and energy-efficient maritime communication multi-objective optimization problem (SEMCMOP) to enhance transmission efficiency, security, and minimize energy consumption. The SEMCMOP is an NP-hard and large-scale optimization problem, making it more complex to solve.

\item \textbf{\textit{Improved Swarm Intelligence Optimization Algorithm:}} We utilize swarm intelligence optimization algorithms for dealing with the complex SEMCMOP. Specifically, we propose an improved multi-objective mayfly algorithm (IMOMA) with chaotic solution initialization and hybrid solution update strategies to optimize the UAVs. The IMOMA can enhance the diversity of initial solutions and update the solutions in different dimensions in a targeted manner.

\item\textbf{\textit{Simulations and Findings:}} The simulation results demonstrate that the CB-based method can achieve more efficient and secure long-distance signal transmission compared to non-CB, single CB and multi-hop approaches. Moreover, comparison results show that the proposed IMOMA outperforms other contrasting swarm intelligence algorithms. In addition, IMOMA is particularly significant in protecting against eavesdropping, improving the security-related objective by up to 43.20$\%$, making it highly suitable for secure maritime communications.
\end{itemize}

\par The rest of this work is organized as follows: Section~\ref{sec:related work} reviews the related work. Section~\ref{sec:models and preliminaries} gives the models and preliminaries. Section~\ref{sec:problem formulation and analysis} formulates the SEMCMOP. Section~\ref{sec:algorithm} presents the algorithm. Section~\ref{sec:simulation results and analysis} illustrates the simulation results. Section~\ref{sec:discussion} supplements the relevant discussion and Section~\ref{sec:conclusion} summarizes the paper.

%

\section{Related Work}
\label{sec:related work}

\par In this section, we review the works associated with relay-assisted maritime wireless communications, maritime communication security strategies, and multi-objective optimization problems. Moreover, we summarize the differences between existing works and current work in Table \ref{table：related-work}.

\begin{table*}[htbp]
\caption{Comparison between related works and this work}
\label{table：related-work}
\newcommand{\tabincell}[2]{\begin{tabular}{@{}#1@{}}#2\end{tabular}} 
\begin{center} 
\renewcommand\arraystretch{1.2}
\begin{tabular}{|c|c|c|c|c|c|c|c|c|}
\hline  \quad & \multicolumn{2}{c|}{Considered scenarios} & \multicolumn{2}{c|}{Security}  & \multicolumn{3}{c|}{Optimization objectives} & Optimization methods \\
\hline Reference & Maritime scenario & \tabincell{c}{UAV-assisted \\ relay} & PLS  & \tabincell{c}{UAV-assisted \\ jamming }  & \tabincell{c}{Signal \\ transmission} &  Security & \tabincell{c}{Energy \\consumption} & \tabincell{c}{Swarm intelligence\\algorithm}\\
\hline \cite{Zeng2023}  & \checkmark & \ding{53} & \ding{53} &\ding{53} & \ding{53}  & \ding{53}  & \ding{53} & \ding{53} \\
\hline \cite{Hu2024} &\checkmark &\ding{53} &\ding{53} & \ding{53} &\ding{53} & \ding{53} & \ding{53} & \ding{53}\\
\hline \cite{Wu2024}  &\checkmark & \ding{53} & \ding{53} &\ding{53} & \ding{53}  & \ding{53}  & \ding{53} & \ding{53}\\
\hline \cite{wang2022unmanned}  & \checkmark & \ding{53} & \ding{53} & \ding{53} & \ding{53} &\ding{53} & \ding{53} & \ding{53}\\ 
\hline \cite{zeng2021joint}  & \checkmark &  \ding{53} & \ding{53} & \ding{53} & \ding{53} &\ding{53} & \ding{53} & \ding{53}\\ 
\hline \cite{Liu2022}  & \checkmark & \checkmark & \ding{53} & \ding{53} & \ding{53} &  \ding{53} & \ding{53} & \ding{53}\\ 
\hline \cite{Qian2023}  & \checkmark & \checkmark & \ding{53} & \ding{53} & \ding{53} &  \ding{53} & \checkmark  & \ding{53} \\  
\hline \cite{aman2023security} &\checkmark & \ding{53} & \ding{53} & \ding{53} & \ding{53} & \checkmark & \ding{53} &\ding{53} \\
\hline \cite{Vangala2024} & \checkmark & \ding{53} & \ding{53} & \ding{53} & \ding{53} & \checkmark & \ding{53} &\ding{53} \\ 
\hline \cite{Ren2023} & \ding{53} & \ding{53}  & \ding{53} & \checkmark & \ding{53} &\checkmark & \ding{53} & \ding{53}\\
\hline \cite{Zhao2018} & \checkmark & \ding{53} & \ding{53} & \checkmark & \ding{53} & \ding{53} & \ding{53} & \ding{53}\\
\hline \cite{Wang2021a} & \ding{53} &  \checkmark  & \checkmark & \checkmark & \checkmark &\checkmark & \ding{53} & \ding{53} \\
\hline \cite{Lu2024} &\checkmark & \ding{53} & \checkmark  & \checkmark  &\ding{53} & \checkmark & \ding{53} &\ding{53}\\
\hline \cite{Yang2024} & \checkmark & \ding{53} & \checkmark & \checkmark & \ding{53} & \checkmark &\ding{53} & \ding{53}\\
\hline \cite{Luo2024} & \checkmark & \checkmark & \ding{53} & \ding{53} & \ding{53} &  \ding{53} & \ding{53} & \ding{53} \\
\hline \cite{Hashim2019} & \checkmark & \ding{53} & \checkmark & \checkmark & \ding{53} & \checkmark &\ding{53} &\checkmark\\
\hline \cite{Qiu2020}  & \ding{53} &\checkmark & \ding{53} & \ding{53} & \ding{53} &\ding{53} & \ding{53} & \checkmark  \\
\hline This work & \checkmark & \checkmark & \checkmark & \checkmark & \checkmark & \checkmark & \checkmark & \checkmark\\
\hline 
\end{tabular}
\end{center}
\end{table*}

\subsection{Relay-assisted Maritime Communications}

\par Maritime communications are vital for vessel navigation and emergency response. Due to the difficulty of deploying equipment, relying on auxiliary tools to enhance network performance is necessary~\cite{Zeng2023}. For example, Hu \textit{et al.}~\cite{Hu2024} proposed a theoretical framework for a low Earth orbit (LEO) satellite-aided shore-to-ship communication network to obtain the end-to-end transmission performance by considering signal transmissions through either a marine link or a space link. Wu \textit{et al.}~\cite{Wu2024} introduced an intelligent spectrum-sharing strategy for satellite maritime networks, enabling satellites to evaluate channel allocation actions to optimize throughput and spectrum efficiency. However, satellite-assisted maritime communications face latency problems over long distances. Moreover, Wang \textit{et al.}~\cite{wang2022unmanned} utilized an unmanned surface vessel (USV) to assist maritime wireless communications and demonstrated the benefits of USV-assisted mobile relaying. Zeng \textit{et al.}~\cite{zeng2021joint} considered a USV-enabled maritime wireless network, where a USV is employed to assist the communications between the terrestrial base station and ships. However, the off-shore propagation conditions are influenced by sea surface reflection and scattering, which lead to multipath effects and deteriorate the quality of the received signals~\cite{liu2021novel}~\cite{Stove2017}. Moreover, the slow mobility of USVs limits their communication coverage and flexibility, and waves can pose safety risks to their operation.

\par In recent years, a UAV has been used as an effective and convenient tool to assist maritime communications due to its flexibility and ease of deployment. For example, Liu \textit{et al.}~\cite{Liu2022Deep} established a two-layer UAV-enabled maritime communication network, which is employed to solve the latency minimization problem for computation and communication. Qian \textit{et al.}~\cite{Qian2023} considered a UAV-assisted maritime Internet of Things (M-IoT) network to improve the workload computation and energy efficiency of offloading transmission. However, UAVs operating at higher altitudes may encounter signal attenuation over long distances, thereby adversely impacting the overall communication performance. In this case, CB can adjust the amplitude and phase of signals through the cooperation of multiple transmitting antennas so that the signals can be superimposed at the receiving end, thus improving signal strength and coverage~\cite{Li2024}. Therefore, based on CB, multiple UAVs form a VAA to enhance the synthetic gain of the signal to achieve long-distance maritime communications.

\subsection{Maritime Communication Security Strategies}

\par Due to the open nature of the maritime channels, security issues need to be taken into account during communications. For instance, Aman \textit{et al.}~\cite{aman2023security} systematically discussed the security needs and solutions of air–water wireless communication networks. Vangala \textit{et al.}~\cite{Vangala2024} proposed a new lightweight authentication protocol by utilizing drone technology in conjunction with the 5G mobile network communications, withstanding various security attacks and maintaining low communication and computation costs. Ren \textit{et al.}~\cite{Ren2023} presented a novel physically unclonable function-based access authentication scheme to achieve mutual authentication and privacy protection in the UAV-aided satellite-terrestrial integration networks. However, the energy of the encryption and decryption methods in the abovementioned works depends on the amount of transmitted data. When the data is larger, computational energy is more immense, making the methods unsuitable for an energy-limited maritime environment. In addition, complex key distribution and management mechanisms increase the complexity of communications.

\par Since the PLS can dynamically adjust the security mechanism based on the channel states, and the dynamic deployment characteristics of UAVs, there have been many studies that considered UAVs for PLS maritime communications. For instance, Wang \textit{et al.}~\cite{Wang2021a} investigated a dual-UAV-enabled secure communication system, in which a UAV sends confidential messages to a mobile user while another cooperative UAV sends artificial noise signals to confuse malicious eavesdroppers, improving a worst-case secrecy rate. Lu \textit{et al.}~\cite{Lu2024} proposed an efficient secure communication scheme for UAV-relay-assisted maritime mobile edge computing (MEC) with a flying eavesdropper, to maximize the secure computing capacity of maritime devices. Liu \textit{et al.}~\cite{Liu2022} proposed a reinforcement learning-based UAV relay policy for maritime communications to resist jamming attacks and decrease the bit-error-rate of the maritime signals. Note that the implementations of the abovementioned works require optimizing the 3D trajectories or power allocation of UAVs. However, the power allocation approach can decrease the communication rates of the target receivers. In addition, the UAVs need to fly from original locations to target locations to improve communication performance, which inevitably increases their flight energy consumption and reduces the corresponding lifetime. Therefore, UAVs can form VAA to send jamming signals to the remote illegitimate eavesdropper via CB, which allows UAVs to achieve secure maritime communications without long-distance flight. Note that CB has limitations, such as increased communication overhead for data sharing and limited support for multiple users. However, it remains a promising solution for enhancing secure maritime communications, offering significant benefits in signal quality and energy efficiency.

\subsection{Multi-objective Optimizations}

The previous approach, which combines multiple objectives into a single one~\cite{Liu2022}~\cite{Wang2019a}, can be effective while it lacks flexibility, making it difficult to quickly evaluate and select the most appropriate trade-offs. In this case, multi-objective optimization methods can be used to address various trade-offs in different scenarios, enabling optimal decision-making. Consequently, the following are multi-objective optimization methods for handling MOP. First, the weighted sum method transforms multiple objectives into a single objective by weighted summing, where different objectives are assigned different weights, and the total value of the objective function is the weighted sum~\cite{Khan2021}~\cite{Lin2024}. However, the method requires predefined weight coefficients and may reduce the solution space. Moreover, when the Pareto front (PF) is non-convex, the weighted sum method may not get the complete set of Pareto optimal solutions. Second, deep reinforcement learning (DRL) is increasingly used to solve MOP by learning policies, which integrates PF approximation to balance multiple conflicting objectives. For instance, Yang \textit{et al.}~\cite{Yang2024} explored a UAV-assisted maritime communication scheme using reconfigurable intelligent surfaces to enhance energy efficiency while defending against jamming attacks, ensuring quality of service. Luo \textit{et al.}~\cite{Luo2024} developed a DRL-optimized method that allows UAVs to predict buoy positions and optimize movement control to enhance beam pointing and maintain stable line-of-sight (LoS) communication for efficient maritime data transmission. However, DRL is more suitable for continuous time-slot problems in real-time decision-making scenarios. When applied to the transient MOP in this paper, it may incur additional computational overhead in the training phase, wasting valuable maritime resources. In addition, DRL can further address the challenge of vessels in continuous motion, requiring UAVs to dynamically adjust their positions to maintain communications, which will be explored in subsequent work.

\par Furthermore, multi-objective swarm intelligence optimization algorithms introduce Pareto dominant to find a set of candidate solutions for MOP. For example, Hashim \textit{et al.}~\cite{Hashim2019} proposed the multi-objective particle swarm optimization to trade off two objectives and defined a set of non-dominated solutions on the Pareto front that gave the optimal compromise solutions. Qiu \textit{et al.}~\cite{Qiu2020} adapted a multi-objective pigeon-inspired optimization algorithm to coordinate UAVs, ensuring stable flight formations in complex environments. However, our problem contains a large number of variables with different boundary values, which is challenging for classical swarm intelligence algorithms. Therefore, we intend to propose a novel swarm intelligence algorithm to handle multiple decision variables in the considered scenario.

\par Different from previous works, this paper utilizes UAVs to achieve remote maritime communications based on CB while noting the security issues of the process. Moreover, a corresponding improved algorithm is proposed for solving the MOP in this scenario.In summary, our approach uniquely extends transmission range, enhances security, and energy efficiency, making a significant contribution to secure maritime communications.

%

\section{Models and preliminaries}
\label{sec:models and preliminaries}

\par In this section, we present the CB-based dual UAV cluster-assisted maritime secure communication system. Then, we give the communication models and energy consumption model of the UAV.

\subsection{System Overview}

\par
\begin{figure}[!t]
\centering
\includegraphics[width=3.5in]{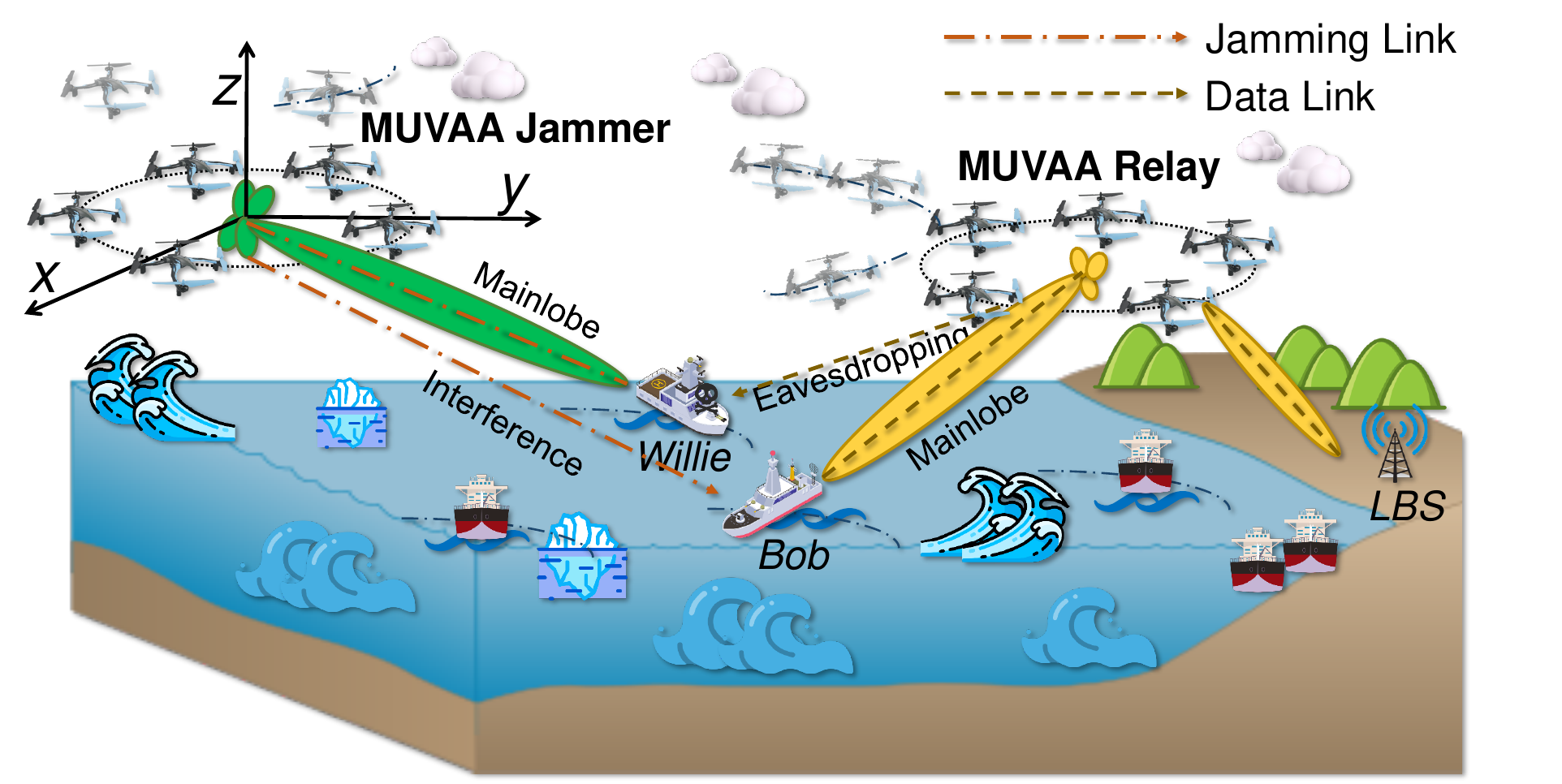}
\caption{A CB-based dual UAV cluster-assisted maritime secure communication system.}
\label{fig:scenario-model}
\end{figure}

 Fig.~\ref{fig:scenario-model} shows the CB-based dual UAV cluster-assisted maritime secure communication system model, which includes a land base station (LBS), a legitimate vessel denoted as Bob, an illegitimate vessel denoted as Willie, and two UAV clusters. Due to infrastructure limitations in the maritime environment and the large vessel Bob cannot move close to the coast, it is challenging for the LBS to communicate with the remote Bob directly. Therefore, a set of rotary-wing UAVs denoted as $\mathcal{U}_{R} $ = $\left \{ 1,2,..., N_{UR}\right \}$ are dispatched as a cluster to receive and forward data signals by the data link, whereas Willie aims to eavesdrop on the link. To restraint Willie, the other set of UAVs marked as $\mathcal{U}_{J} $ = $\left \{ 1,2,..., N_{UJ}\right \}$ sends jamming signals to Willie. In this work, the specific shipping-lanes can be obtained to determine the directions and locations of vessels in advance. Moreover, the UAV is assumed to be fitted with a single omnidirectional antenna and global positioning system (GPS), and the information of illegitimate vessels can be detected by optical cameras or synthetic aperture radar installed on the UAV.

\par The process begins with LBS sending data signals to the dispatched $\mathcal{U}_{R}$ UAVs by the ground-to-air (G2A) data link. Next, the UAV cluster forms the MUVAA relay and forwards data signals to Bob by the air-to-sea (A2S) data link. Then, $\mathcal{U}_{J}$ UAVs can depart from their original hovering positions, and form an MUVAA jammer to send jamming signals to Willie by the A2S jamming link. We consider that the UAVs in the same virtual antenna array are synchronized in terms of the carrier frequency, initial phase, and time~\cite{Mohanti2022}, and these UAVs can achieve data sharing by using the method in~\cite{Feng2013a}. Note that we analyze the operations of UAVs in a specific scenario. Specifically, Bob and Willie are located at fixed positions when the UAVs send signals to them. This allows us to derive a clear understanding of the performance of the system under specific conditions and lays the foundation for future studies of dynamic factors.

\par In this process, the LBS as the central node of data signals can efficiently get the channel state information (CSI) of objects through centralized control, feedback mechanisms, and channel estimation techniques~\cite{Guo2022}. Moreover, we consider that the UAVs obtain the quantified version of the CSI through the approaches in ~\cite{Ahmad2022}. The balance between the CSI code rates and the quantization errors needs to be optimized, as the lower code rate can reduce the description of the CSI accuracy and increase the errors. To demonstrate the general applicability of this work, we employ the 3D Cartesian coordinate system, in which the positions of Bob, Willie, LBS, the $m$th UAV in the MUVAA relay, and $n$th UAV in the MUVAA jammer are denoted as ($x^{B}$, $y^{B}$, $z^{B}$), ($x^{W}$, $y^{W}$, $z^{W}$), ($x^{L}$, $y^{L}$, $z^{L}$), ($x_{m}^{Ur}$, $y_{m}^{Ur}$, $z_{m}^{Ur}$), and ($x_{n}^{Uj}$, $y_{n}^{Uj}$, $z_{n}^{Uj}$), respectively. Subsequently, we give the key model associated with communications.

\subsection{Communication Models}

\par In our considered system, there are three types of communication links: (1) the G2A data link to send data signals from the LBS to MUVAA relay, (2) the A2S data link to forward data signals between the MUVAA relay and Bob, which might be eavesdropped by Willie, (3) the A2S jamming link from the MUVAA jammer that is used to send jamming signals to Willie and might interfere with Bob. Next, we elaborate on the three links.

\subsubsection{G2A Data Link from the LBS to MUVAA Relay}

\par In the considered system, a CB-based UAV relay can forward data signals from an LBS to Bob, and the specific process is as follows. First, the LBS utilizes CB to send data signals to UAV in $\mathcal{U}_{R}$, which immediately caches or forwards the signals based on channel conditions or node ranges. Specifically, the LBS is usually configured with a large number of antennas and can perform channel estimation through massive multiple-input multiple-output (MIMO)~\cite{Wang2024}. In addition, the LBS has powerful computation and processing capabilities to obtain accurate CSI of the UAVs. Therefore, the LBS can calculate the weights of beam patterns based on CSI to focus on the UAV direction, thus efficiently sending signals over longer distances. Then, the UAV broadcasts the received data signals to all UAVs in $\mathcal{U}_{R}$. Due to the high altitude of the UAVs, the airborne transmission follows the LoS channel conditions~\cite{Li2024}. Finally, the UAVs form the MUVAA relay based on the assigned weights for coordinated transmission.

\par We can accomplish the abovementioned process by satisfying the information dissemination constraint in the following ways. \textbf{\textit{First}}, the airborne signal sharing process among UAVs in the MUVAA relay usually has a small transmission distance and good channel conditions, which achieve high broadcast rates of airborne UAVs. Moreover, the signal sharing process can be implemented by using many low-cost and efficient methods~\cite{Feng2013a}. Therefore, the transmission rates of the CB-based relay system are not constrained by the airborne signal sharing phase. \textbf{\textit{Second}}, our proposed signal relaying method can be offline, i.e., the data has been uploaded to the UAV. At this point, the transmission rates from the LBS to the UAV in the MUVAA relay can be reasonably omitted. Moreover, the LBS can dynamically adjust the uplink transmission rates depending on the signal relay rates of the UAV. Therefore, the information transfer performed in the G2A data link is reliable. \textbf{\textit{Finally}}, the UAVs usually have a storage device with some caching capability. Therefore, when the signal rates of the G2A data link are higher than that of the A2S data link, the UAVs can first cache some of the data and then send it to the legitimate vessel in unison~\cite{Sun2022}. In summary, in our considered system, the LBS can conduct spectrum and power allocation, and it has adequate transmission power~\cite{Gong2014}, which means that the LBS can automatically adapt the G2A data link transmission rates by the A2S data link.

\subsubsection{A2S Data Link from MUVAA Relay to Bob}

\par Mathematically, we use the array factor to measure the strength of the data signals in different directions for MUVAA relay~\cite{Mozaffari2019}, which is represented as follows:
\begin{equation}
\begin{split}
AF_{r}&(\theta,\phi) = \\&\sum_{m=1}^{N_{UR}}I_{m}^{Ur} e^{\iota[k_{c}(x_{m}^{Ur}\sin\theta \cos\phi+ y_{m}^{Ur}\sin\theta\sin\phi+z_{m}^{Ur}\cos\theta)]},
\label{con:AF1}
\end{split}
\end{equation}
\noindent where $I_{m}^{Ur}$ denotes the excitation current weight of the $m$th UAV in the MUVAA relay, $\theta \in[0,\pi]$ and $\phi\in[-\pi,\pi]$ denote the elevation and azimuth angles under the A2S data link, respectively. In addition, $\iota$ is imaginary units, $k_{c} = 2\pi /\lambda$ and $\lambda$ denotes the wavelength.

\par Then, the antenna gain from the UAVs in the MUVAA relay to the vessel is denoted by
\begin{equation}
G_{v} (\mathbb{P}_{r}) = \frac{4\pi|AF_{r}(\theta_{v},\phi_{v})|^{2} \omega(\theta_{v},\phi_{v})^{2}}{\int_{0}^{2\pi }\int_{0}^{\pi} |AF_{r}(\theta ,\phi)|^{2} \omega(\theta,\phi)^{2}\sin\theta d\theta d\phi}\eta,
\label{con:G_R}
\end{equation}
\noindent where $\mathbb{P}_{r}=\left \{ \mathbb{X}_{r},\mathbb{Y}_{r},\mathbb{Z}_{r}\right\}$ denotes the set of UAV positions in the MUVAA relay, which is one of the decision variables of the system. The definitions of this variable and other relevant variables in this paper are detailed in Table \ref{table:variables}. Moreover, $(\theta_{v},\phi_{v})$ and $\omega(\theta,\phi)$ denote the direction towards a vessel (either Bob or Willie) and magnitude of the far-field beam pattern of a UAV under the A2S data link, respectively, and $\eta \in [0,1]$ denotes the antenna array efficiency~\cite{Sun2022}.

\par Then, during the A2S transmission, the antenna heights of UAVs are significantly greater than those of vessels. Thus, the path loss from MUVAA relay to the vessel can be expressed as follows~\cite{Wang2021}:
\begin{equation}
\begin{split}
    PL(\mathbb{P}_{r})[dB]=&\frac{A_U}{1+\alpha_{a}e^{-\alpha_{b}(\theta-\alpha_{a})}}+20\log_{10}^{d}+C_{r}\\& +20\log_{10}^{({4\pi f_{c}}/{300})},
    \end{split}
    \label{PL_R}
\end{equation}
\noindent where $d={\sqrt{(x_{r}-x_{v})^{2}+(y_{r}-y_{v})^2+(z_{r}-z_{v})^2}}$, and $\theta=({180}/{\pi}) \arcsin({z_{r}}/d)$, wherein $x_{r}=\mathbb{E} (\mathbb{X}_{r}), y_{r}=\mathbb{E}(\mathbb{Y}_{r}), z_{r}=\mathbb{E} (\mathbb{Z}_{r})$. $\mathbb{E}(\cdot)$ is a mean operator that is used to calculate the average value of each row of a matrix, and $(x_{v}, y_{v}, z_{v})$ is the 3D location at a vessel (either Bob or Willie). Moreover, $f_{c}$ is the carrier frequency in MHz, and $A_{U}$, $C_{r}$, $\alpha_{a}$ and $\alpha_{b}$ are environment-related constant parameters in dB.

\subsubsection{A2S Jamming Link from MUVAA Jammer to Willie} 

\par Similarly, the array factor is used to evaluate the strength of jamming signals of the MUVAA jammer, which is given by
\begin{equation}
\begin{split}
A&F_{j}(\theta^{'},\phi^{'}) = \\&\sum_{n=1}^{N_{UJ}}I_{n}^{Uj} e^{\iota[k_{c}(x_{n}^{Uj}\sin\theta^{'}\cos\phi^{'}+ y_{n}^{Uj}\sin\theta^{'}\sin\phi^{'}+z_{n}^{Uj}\cos\theta^{'})]},
\label{con:AF2}
\end{split}
\end{equation}
\noindent where $I_{n}^{Uj}$ denotes the excitation current weight of the $n$th UAV in the MUVAA jammer, and $\theta^{'} \in[0,\pi]$ and $\phi^{'}\in[-\pi,\pi]$ denote the elevation and azimuth angles under the jamming link, respectively.

\par Correspondingly, the antenna gain from the UAVs in the MUVAA jammer to the vessel is as follows:
\begin{equation}
G^{'}_{v} (\mathbb{P}_{j}) = \frac{4\pi|AF_{j}(\theta_{v}^{'},\phi_{v}^{'})|^{2} \omega(\theta _{v}^{'},\phi_{v}^{'})^{2}}{\int_{0}^{2\pi }\int_{0}^{\pi} |AF_{j}(\theta^{'} ,\phi^{'})|^{2} \omega(\theta^{'},\phi^{'})^{2}\sin\theta^{'} d\theta^{'} d\phi^{'}}\eta,
\label{con:G_J}
\end{equation}
\noindent where $\mathbb{P}_{j}=\left \{ \mathbb{X}_{j},\mathbb{Y}_{j},\mathbb{Z}_{j}\right\}$ denotes the set of UAV positions in the MUVAA jammer, which is shown in Table \ref{table:variables}. Moreover, $(\theta_{v}^{'},\phi_{v}^{'})$ is the direction to a vessel (either Bob or Willie) under the jamming link, and $\omega(\theta^{'},\phi^{'})$ denotes the magnitude of the far-field beam pattern of a UAV under the jamming link. 

\par Then, from the MUVAA jammer, the transmission path loss towards the vessel is calculated by
\begin{equation}
\begin{split}
    PL^{'}(\mathbb{P}_{j})[dB]=&\frac{\eta _{LOS} -\eta _{NLOS}}{1+\alpha_{a}e^{-\alpha_{b}(\theta^{'}-\alpha_{a})}}+\eta _{NLOS}\\& +20(\log_{10}^{({4\pi f_{c}}/{300})}+\log_{10}^{d^{'}}),
    \end{split}
    \label{PL_j}
\end{equation}
\noindent where $d^{'} = {\sqrt{(x_{j}-x_{v})^{2}+(y_{j}-y_{v})^2+(z_{j}-z_{v})^2}}$ is the distance between the center of MUVAA jammer and the vessel (either Bob or
Willie), wherein $x_{j}=\mathbb{E} (\mathbb{X}_{j}), y_{j}=\mathbb{E}(\mathbb{Y}_{j}), z_{j}=\mathbb{E} (\mathbb{Z}_{j})$. Moreover, $\theta^{'}=({180}/{\pi}) \arcsin( {z_{j}}/d^{'})$, $\eta _{LOS}$ and $\eta_{NLOS} $ are the corresponding parameters shown in Table~\ref{table:table1}.

\par Signal-to-interference-plus-noise ratio (SINR) is a key metric to measure signal quality in wireless communications. Specifically, SINR reflects the channel condition by the ratio of the useful signal strength to the sum of interference and noise. Therefore, based on the A2S data link and A2S jamming link, the obtainable SINR of Bob is expressed as:
\begin{equation}
\gamma_{Bob}=\frac{P_{UR}N_{UR}G_{B}PL_{B}}{P_{UJ}N_{UJ}G_{B}^{'}PL_{B}^{'}+\sigma^{2}},
    \label{sinr_Bob_sim}
\end{equation}
\noindent where $P_{UR}$ and $P_{UJ}$ are the transmission powers of UAV in the MUVAA relay and MUVAA jammer, $N_{UR}$ and $N_{UJ}$ are the numbers of UAVs in the MUVAA relay and MUVAA jammer, respectively. Moreover, $G_{B}$ and $G_{B}^{'}$ can be obtained by replacing Eqs. (\ref{con:G_R}) and  (\ref{con:G_J}) with $(\theta_{v}, \phi_{v})=(\theta_{B},\phi_{B})$, $(\theta_{v}^{'}, \phi_{v}^{'})=(\theta_{B}^{'},\phi_{B}^{'})$, $PL_{B}$ and $PL_{B}^{'}$ are required according to the (\ref{PL_R}) and (\ref{PL_j}) with  $d = d_{B}$ and $d^{'}$ = $d^{'}_{B}$, respectively. In addition, $\sigma^{2}$ is the additive white Gaussian noise.

\par Correspondingly, we set $(\theta_{v},\phi_{v})=(\theta_{W},\phi_{W})$, $d = d_{W}$, $(\theta_{v}^{'},\phi_{v}^{'})=(\theta_{W}^{'},\phi_{W}^{'})$, and $d^{'}$ = $d^{'}_{W}$ to replace Eqs. (\ref{con:G_R}), (\ref{PL_R}), (\ref{con:G_J}) and (\ref{PL_j}), the obtainable SINR of Willie can be expressed as:
\begin{equation}
    \gamma_{Willie}=\frac{P_{UR}N_{UR}G_{W}PL_{W}}{P_{UJ}N_{UJ}G_{W}^{'}PL_{W}^{'}+\sigma^{2}}.
    \label{sinr_Willie_sim}
\end{equation}

\par As aforementioned, the positions of the vessels are not adjustable, and the 3D positions and excitation current weights of UAVs are the key decision variables to affect maritime communications effectiveness. Moreover, the process of regulating the 3D positions of the UAVs consumes their energy. Next, we introduce the moving energy consumption model for the UAV.
 
\begin{table*}[!t]
\renewcommand{\arraystretch}{1.3}
\newcommand{\tabincell}[2]{\begin{tabular}{@{}#1@{}}#2\end{tabular}}
\caption{The definition of the variables}
\label{table:variables}
\begin{center} 
\begin{tabular}{c|c|c}
\hline
\textbf{ \tabincell{c}{Variable \\expression}} & \textbf{Variable elements} & \textbf{Variable Declarations}\\
\hline $\mathbb{P}_{r}$ & \tabincell{c}{
$\left \{\mathbb{P}_{r}=\mathbb{X}_{r},\mathbb{Y}_{r}, \mathbb{Z}_{r}\right \} $,\\
$\left \{\mathcal{P}_{m}^{Ur} = (x_{m}^{Ur},y_{m}^{Ur},z_{m}^{Ur})| m \in \mathcal{U}_{R} \right \} $} & \tabincell{c}{ $\mathbb{P}_{r}$ represents the relay set including positions of all UAVs in the MUVAA relay, \\$\mathbb{X}_{r}$ and $\mathbb{Y}_{r}$ are the horizontal positions of UAVs, $\mathbb{Z}_{r}$ is the vertical positions of UAVs,\\$\mathcal{P}_{m}^{Ur}$ represents the position of the $m$th UAV in the MUVAA relay.}\\
\hline $\mathbb{P}_{j} $ & \tabincell{c}{
$\left \{\mathbb{P}_{j}=\mathbb{X}_{j},\mathbb{Y}_{j}, \mathbb{Z}_{j}\right \} $,\\ $\left \{\mathcal{P}_{n}^{Uj} = (x_{n}^{Uj},y_{n}^{Uj},z_{n}^{Uj})| n\in \mathcal{U}_{J} \right \} $} & \tabincell{c}{$\mathbb{P}_{j}$ denotes the set of positions of all UAVs in the MUVAA jammer, \\$\mathbb{X}_{j}$ and $\mathbb{Y}_{j}$ are the horizontal positions of UAVs, $\mathbb{Z}_{j}$ is the vertical positions of UAVs,\\$\mathcal{P}_{n}^{Uj}$ represents the position of the $n$th UAV in the MUVAA jammer.}\\
\hline $\mathbb{I}_{r} $ & $\left \{ I_{m}^{Ur}|m\in \mathcal{U}_{R} \right \} $ & \tabincell{c}{$\mathbb{I}_{r} $ is the set of excitation current weights of all UAVs in the MUVAA relay, \\$I_{m}^{Ur}$ is the excitation current weight of the $m$th in the MUVAA relay.}\\
\hline $\mathbb{I}_{j} $ & $\left \{ I_{n}^{Uj}|n\in \mathcal{U}_{J} \right \} $ & \tabincell{c}{ $\mathbb{I}_{j} $ is the set of excitation current weights of all UAVs in the MUVAA jammer, \\$I_{n}^{Uj}$ is the excitation current weight of the $n$th in the MUVAA jammer.}\\
\hline
\end{tabular}
\end{center}
\end{table*}

\subsection{Energy Consumption Model of the UAV}

\par In general, the communication and propulsion energy consumption compose the total flight energy consumption of UAVs. However, the value of communication energy consumption is minimal, and it is often neglected in the calculation~\cite{Li2021}. Thus, when a rotary-wing UAV flies horizontally in 2D, the propulsion power consumption can be calculated by
\begin{equation}
\begin{split}
P(v) = &P_{I}(\sqrt{1+\frac{v^{4}}{4v_{m}^{4}}} -\frac{v^{2} }{2v_{m}^{2}}) ^{\frac{1}{2}} +P_{B}(1+\frac{3v^{2}}{v_{t}^{2}}) \\& + \frac{1}{2}d_{f}s_{r}\rho_{a}a_{r}v^{3},
\label{p_v}
\end{split}
\end{equation}
\noindent where $v$ is the velocity of the UAV, and $P_{I}$ and $P_{B}$ represent the induced power and blade profile power in the hovering conditions, respectively. $v_{m}$ is the mean rotor induced velocity in hovering, $v_{t}$ is the tip speed of the rotor blade, and $d_{f}$, $s_{r}$, $\rho_{a}$ and $a_{r}$ represent the fuselage drag ratio, rotor solidity, air density, and rotor disc area, respectively. 

\par Note that the additional energy consumption of UAVs from acceleration and deceleration during horizontal flight is negligible, as it takes up only a small fraction of the entire running time of UAVs. Therefore, based on the propulsion energy consumption, movement and gravity energy consumption in the case of ascent and descent with time, the energy consumption of 3D trajectory of UAV using the heuristic closed-form approximation is expressed by~\cite{Zeng2019}
\begin{equation}
\begin{split}
E\left(T\right)\approx&\int_{0}^{T}P\left(v\left(t\right )\right) dt+\frac{1}{2}m_{U}\left (v\left(T\right )^{2}-v\left(0 \right)^{2} \right)\\&+m_{U}g\left(h\left(T\right)-h\left(0\right)\right ),
\end{split}
\label{E_t}
\end{equation}
\noindent where $v(t)$ is the instantaneous velocity of the UAV at time $t$, and $T$ is the time duration of the UAV flight. Moreover, $m_{U}$ and $g$ are the aircraft mass of a UAV and gravitational acceleration, respectively.

\par According to the energy consumption model, we can summarize that the energy consumption of the UAVs is primarily related to their positions. Therefore, the positions of UAVs in the MUVAA relay and MUVAA jammer have a critical influence on communication effectiveness.

%

\section{Problem formulation and analysis} \label{sec:problem formulation and analysis}

\par In this section, we specify the problem of the considered system. Then, we propose the optimization objectives and formulate the SEMCMOP. Next, the problem is analyzed.

\subsection{Problem Statement}

\par The main objective of this paper is to achieve remote maritime communications and ensure security while saving energy consumption of the UAV. On the one hand, since the LBS is far from Bob, making direct communication challenging, we utilize CB for the remote transmission. Specifically, in a maritime square monitoring area denoted as $A_{sr}$, $N_{UR}$ UAVs form the MUVAA relay and forward data signals to Bob directly. As the vessel moves along its fixed trajectory, the transmission performance of the data signals mainly depends on the beam pattern of the MUVAA relay. To improve the transmission efficiency, the beam patterns can be optimized to point towards Bob to obtain more directional signals. On the other hand, to enhance the communication security, $N_{UJ}$ UAVs forming MUVAA jammer interfere with Willie receiving signals, which may have an impact on Bob. In this case, we optimize the beam patterns of the MUVAA jammer to emit stronger jamming signals toward Willie. As mentioned above, the 3D positions and excitation current weights of UAVs in the VAA jointly determine the beam patterns.



\par According to the abovementioned description, the relevant decision variables to be jointly optimized are as follows: \textit{(i) $ \mathbb{P}_{r} $} denotes the 3D position set of UAVs in the MUVAA relay. \textit{(ii)} $\mathbb{P}_{j} $ denotes the 3D position set of UAVs in the MUVAA jammer. \textit{(iii)} $\mathbb{I}_{r} $ is the excitation current weights set of UAVs in the MUVAA relay. \textit{(iv)} $\mathbb{I}_{j} $ is the excitation current weights set of UAVs in the MUVAA jammer. Note that the optimization variables are specified in the Table \ref{table:variables}.

\subsection{Problem Formulation}

\par In the CB-based dual UAV cluster-assisted maritime secure communication system, we simultaneously reflect on the optimization objectives as follows.

\par \textbf{\textit{Optimization Objective 1:}} To enhance the reliability of legitimate maritime communications, our first optimization objective is to maximize the obtainable SINR value of Bob, and it can be achieved by jointly optimizing the 3D positions and excitation current weights of the UAVs in the MUVAA relay and MUVAA jammer. Thus, the first optimization objective can be given as follows:
\begin{equation}
f_{1}(\mathbb{P}_{r}, \mathbb{I}_{r}, \mathbb{P}_{j}, \mathbb{I}_{j}) = \gamma_{Bob}.
\label{f1}
\end{equation}

\par\textbf{\textit{Optimization Objective 2:}} To maintain the security of legitimate maritime communications, and reduce the risk of eavesdropping on data signals, the second optimization objective is to minimize the available SINR of Willie, which can be expressed as follows:
\begin{equation}
f_{2} (\mathbb{P}_{r}, \mathbb{I}_{r} , \mathbb{P}_{j},\mathbb{I}_{j})= \gamma_{Willie}.
\label{f2}
\end{equation}
 
\par\textit{\textbf{Optimization Objective 3:}} In our designed system, both the MUVAA relay and MUVAA jammer need to move continuously in order to achieve the aforementioned two optimization objectives. Therefore, the third optimization objective is to minimize the total flight energy consumption of UAVs, which is expressed as follows:
\begin{equation}
f_{3}(\mathbb{P}_{r},\mathbb{P}_{j}) = \sum_{m=1}^{N_{Ur}} E_{m}+\sum_{n=1}^{N_{Uj}} E_{n},
\label{f3}
\end{equation}
\noindent where $E_{m}$ and $E_{n}$ represent the flight energy consumption of the $m$th UAV in the MUVAA relay and the $n$th UAV in the MUVAA jammer, respectively. 

\par Note that the three optimization objectives depend on the same decision variables, which means that optimizing one objective can affect others. Moreover, in the process of using UAVs to implement maritime secure communications, the UAVs in the MUVAA relay and MUVAA jammer need to adjust their positions to optimize data transmission rates and regulate the effect of jamming signals. However, the process of constant movements of the UAVs also causes additional energy consumption, which conflicts with our objective of minimizing the total flight energy consumption of UAVs. In addition, according to Eq. (\ref{p_v}), higher UAV speeds increase energy consumption, while slower speeds result in longer hovering durations and increased energy consumption. Therefore, the considered optimization objectives are in conflict with each other and need to be balanced. As such, multi-objective optimization allows for specific trade-offs based on different scenarios, providing flexibility in decision-making.

\par According to the aforementioned three optimization objectives, the SEMCMOP can be formulated as follows:
\begin{subequations}
\label{con:mop}
\begin{align}
\min _{\left \{ \mathbb{P}_{r}, \mathbb{I}_{r},\mathbb{P}_{j},\mathbb{I}_{j} \right \} } \ &F=\left \{ -f_{1},f_{2},f_{3} \right \}, \label{con:mopZa}\\
\mbox{s.t.}\ &0\le I_{m}^{Ur} \le 1, \forall m\in \mathcal{U}_{R},\label{con:mopZb}\\  
&0\le I_{n}^{Uj} \le 1, \forall n\in \mathcal{U}_{J},\label{con:mopZc}\\ 
&\mathcal{P}_{m}^{Ur}\in A_{sr}, \forall m\in \mathcal{U}_{R},\label{con:mopZd} \\
&\mathcal{P}_{n}^{Uj}\in {A}_{sj}, \forall n\in \mathcal{U}_{J},\label{con:mopZe} \\
&DR_{m_{1},m_{2}} \ge D_{min}, \forall m_{1},m_{2}\in \mathcal{U}_{R},\label{con:mopZf}\\
&DJ_{n_{1},n_{2}} \ge D_{min}, \forall n_{1},n_{2}\in \mathcal{U}_{J},\label{con:mopZg}
\end{align}
\end{subequations}
\noindent where $I_{m}^{Ur}$, $I_{n}^{Uj}$, $\mathcal{P}_{m}^{Ur}$ and $\mathcal{P}_{n}^{Uj}$ are the variables associated with UAVs in the relay and jamming sets, respectively, which are displayed in Table \ref{table:variables}. Moreover, $A_{sr}$ and $A_{sj}$ are the 3D coordinates of flight range areas of UAVs in the relay set and jammer set, respectively. In addition, $DR_{m_{1},m_{2}}$ denotes the distance between the $m_{1}$th UAV and $m_{2}$th UAV in the relay set, $DJ_{n_{1},n_{2}}$ denotes the distance between the $n_{1}$th UAV and $n_{2}$th UAV in the jammer set, and $D_{min}$ is the minimum distance between two neighboring UAVs to avoid collision.

\subsection{Problem Analysis}

\par Next, we analyze the formulated SEMCMOP.

\begin{itemize}
\item \textit{The formulated SEMCMOP is NP-hard:} The $f_{3}$ is shown in Eq. (\ref{f3}), the minimization of $f_{3}$ is a continuous optimization problem. For facilitating the analysis, we transform the continuous problem into a discrete problem, such as the solutions of the $x$-coordinate of UAVs in the MUVAA relay ($X_{m}^{Ur}$) are chosen from a set with finite factors. Next, we will verify that the transformed problem is a combinatorial optimization problem.

\par The goal of a combinatorial optimization problem is to identify the optimal subset from a finite universal set that meets specific criteria to achieve the best solutions, and it can be described using three parameters ($F, G$, and $D$). During the minimization of $f_{3}$, where $F$ is the cost function (Eq. (\ref{f3})), $G$ represents the feasible solution region and is a set of constraint functions (Eqs. (\ref{con:mopZd})-(\ref{con:mopZg})), $D$ is the domain of solutions.

\par The transformed version of $f_{3}$ can be regarded as a combinatorial optimization problem which is NP-hard~\cite{Blum2011}. Therefore, the transformed $f_{3}$ is NP-hard, and the initial $f_{3}$ is NP-hard. In addition, the optimization problems related to SINR model ($f_{1}$ and $f_{2}$) are usually NP-hard~\cite{Andrews2009}. Due to $f_{1}$, $f_{2}$, and $f_{3}$ are NP-hard, the originally formulated SEMCMOP is NP-hard. In this case, our current task is to develop an effective algorithm to solve this problem.
\end{itemize}

\begin{itemize}
    \item \textit{The  SEMCMOP is a large-scale optimization problem:} The solution space of the formulated SEMCMOP consists of the 3D positions and excitation current weights of UAVs in the MUVAA relay ($\mathbb{X}_{r},\mathbb{Y}_{r}, \mathbb{Z}_{r}, \mathbb{I}_{r}$) and MUVAA jammer ($\mathbb{X}_{r},\mathbb{Y}_{r}, \mathbb{Z}_{r},\mathbb{I}_{r}$). Thus, the solution dimensions to be processed are ($4\times N_{UR} + 4 \times N_{UJ}$). As the number of UAVs increases, the solution space of the SEMCMOP expands accordingly. Therefore, the formulated SEMCMOP is a large-scale optimization problem.
\end{itemize}

\par Since the formulated SEMCMOP is NP-hard and its complexity increases significantly as the network size increases, it is difficult to find a deterministic algorithm to solve it efficiently. Moreover, due to the SEMCMOP involving a large number of decision variables, and complex constraints and trade-offs among the objectives, weighted sum methods have challenges in facing objective weight setting. In addition, DRL may face convergence difficulties and resource-wasting issues, and it is better suited to address the challenges posed by the dynamic movement of vessels in real-world scenarios~\cite{Sun2023}. In contrast, the multi-objective swarm intelligence optimization algorithms can utilize Pareto dominance to find a set of near-optimal solutions in a short period and choose a suitable final solution from the set according to the requirements of the scenario. Therefore, considering the complexity of SEMCMOP and the limitations of UAV hardware conditions, we propose a novel swarm intelligence optimization algorithm to control the decision variables of SEMCMOP. Additionally, the proposed algorithm runs on Raspberry Pi, which is located on the UAV as an edge computing node, processes the data locally, and uploads the crucial results. It improves the efficiency of data processing and eases the burden on the UAV, thus optimizing the overall resources.

%

\section{The Proposed Algorithm} 
\label{sec:algorithm}

\par In this section, we first introduce the outline of the conventional multi-objective mayfly algorithm (MOMA). Then, the IMOMA is proposed to handle the formulated SEMCMOP.

\begin{figure*}[!t]
\centering
\includegraphics[width=6.5in]{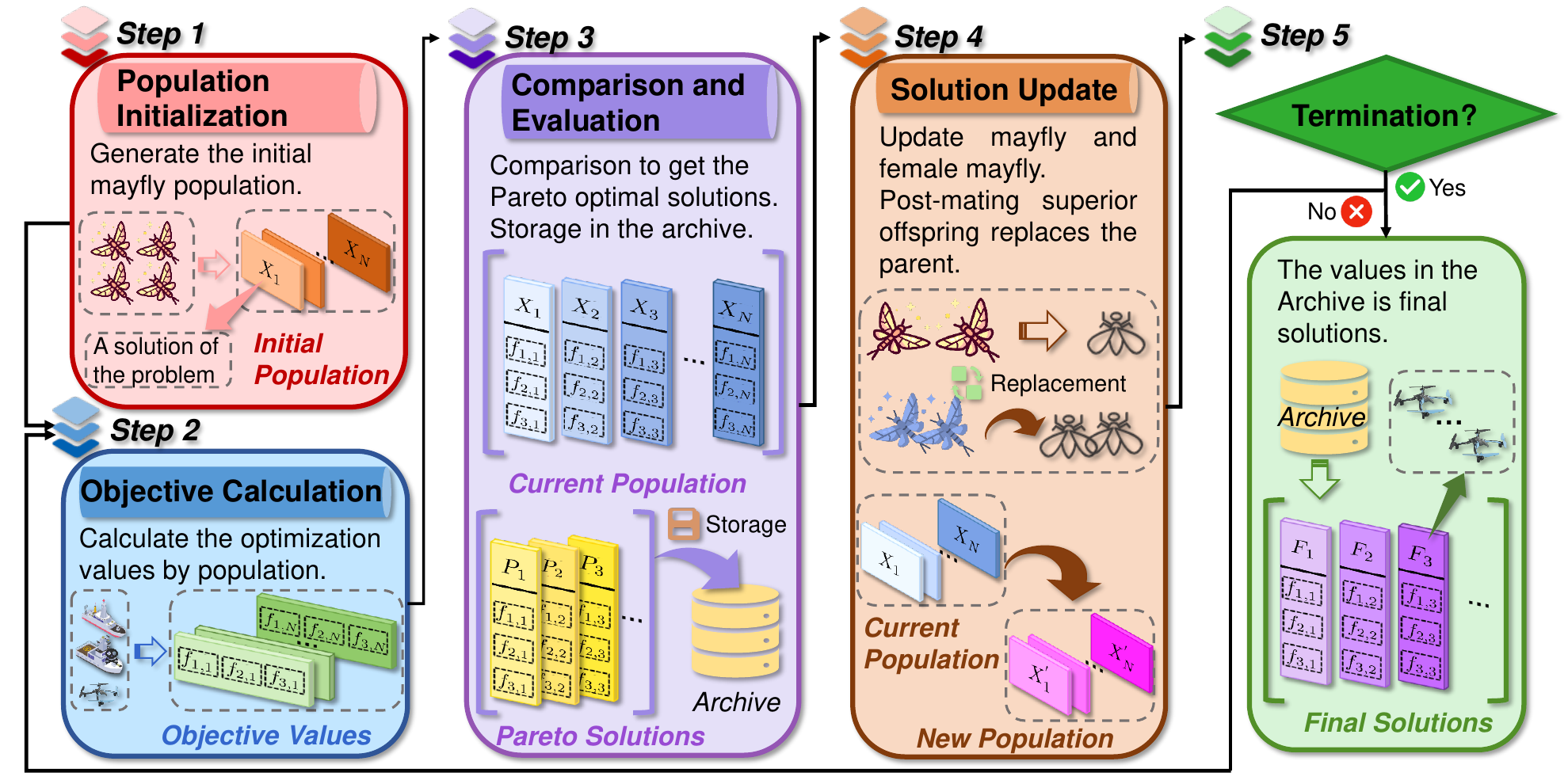}
\caption{Evolutionary outline based on the MOMA.}
\label{fig:outline}
\end{figure*}

\subsection{Outline the Conventional MOMA}

\par First, we describe the advantages of MOMA in dealing with formulated SEMCMOP. Then, we give the overall process of MOMA.

\subsubsection{Advantages of MOMA}

\par Swarm intelligence optimization algorithms have advantages such as parallel search capability and flexible adjustment of solutions, which makes them show efficient performance in dealing with complex MOP~\cite{Yasear2021}. The MOMA is a newly proposed swarm intelligence optimization algorithm and it can find the optimal solutions by simulating the evolution of mayflies and has been applied in practical engineering problems~\cite{Lei2022}. In detail, MOMA has the following advantages. First, male mayflies initially congregate and perform synchronized flights over water to attract females. In response, female mayflies approach the swarm for mating. This process is more effective in balancing exploration and exploitation~\cite{Zhou2022}. Then, after mating, the female mayflies produce offspring, of which only the healthier ones can survive after hatching. If the offspring demonstrates superior fitness, it will displace the weaker parent in the population. This process maintains population diversity and avoids over convergence of the population to the local optimum. More importantly, MOMA is relatively simple and has low computational complexity, and its core mechanisms (mating, displacement) can be executed with smaller computational resources, which is advantageous for large-scale optimization problems
~\cite{Zervoudakis2020}. Therefore, the above features of MOMA make it an ideal choice for handling with the formulated SEMCMOP.

\subsubsection{MOMA Process} The cycling mechanism of MOMA makes it pass on stronger traits to offspring and improve the overall fitness of the population~\cite{Zervoudakis2020}. In this case, the movement of the mayfly can be defined by
\begin{equation}
\label{con:MOMA}
    X_{i}^{t+1}=X_{i}^{t}+v_{i}^{t+1} 
\end{equation}
\noindent where $X_{i}^{t}$ and $X_{i}^{t+1}$ are the current $i$th position of mayfly in the search space at time step $t$ and $t+1$, $v_{i}^{t+1}$ is the velocity of mayfly for changing its position. Moreover, the increasing velocity of male mayfly, female mayfly and offspring is computed differently, which depends on the current different personal best position, global best position, attraction constants, and other corresponding parameters.

\par The position of each mayfly ($X_{i}$) in the search space denotes a prospective solution to the optimization problem. Fig. \ref{fig:outline} shows the outline of MOMA, and the details are described as follows.

\par \textbf{(1) Population Initialization}: Generate the initial population ($X_{N}$) of male and female mayflies at random containing potential solutions to the optimization problem, where $N$ is the population size.

\par \textbf{(2) Objective Calculation}: Calculate the values of the optimization objectives by using the candidate solutions according to Eqs. (\ref{f1})-(\ref{f3}).

\par \textbf{(3) Comparison and Evaluation}: Compare solutions with optimization objective values by Pareto sorting, and obtain the non-dominated Pareto solutions. Then, store them to the archive.

\par \textbf{(4) Solution Update}: Update the mayfly by Eq. (\ref{con:MOMA}) according to the corresponding principle. The male and female mayflies move and mate, and the post-mating superior offspring replaces the poor parent. In turn, the new population replaces the current population.

\par \textbf{(5) Termination or Loop}: Determine if the termination condition is reached. If it is reached, the solutions in the archive are the final solutions; if not, return to \textbf{(2)} for a loop.

\subsection{IMOMA}

\par Since the formulated SEMCMOP has been proven to be an NP-hard and large-scale optimization problem, with optimization objectives need to be balanced, the conventional MOMA may encounter the following challenges in dealing with the problem. 

\begin{itemize}
    \item \textit{Poor Initial Solutions:} The conventional MOMA generates random initial solutions, which can reduce the diversity of solutions and blind the search directions. In addition, due to the solution space of the SEMCMOP shown in Eq. (\ref{con:mop}) is more extensive, poor-quality initial solutions are more prone to falling into local optima~\cite{Kazimipour2014}. Therefore, enhancing the quality of initial solutions is critical.
  \item \textit{Non-uniform Search Space:} The solutions can be continuously updated by the conventional MOMA. However, the values of the optimization objectives are jointly determined by two solution sets $Xr$ and $Xj$, which have different sizes. Moreover, the upper and lower boundaries of each set of solutions are different, which indicates that different dimensions of the algorithm are assigned distinct values. Note that the elements can impact optimal performance, and search strategy may be biased towards certain dimensions, resulting in other dimensions being ignored, which in turn produces the non-uniform solution space and affects overall optimization performance~\cite{Sun2021a}. Therefore, different solution spaces ($Xr$ and $Xj$) and different dimensions of the solutions need to be cautiously considered about developing corresponding update strategies. 
\end{itemize}

\par Therefore, we propose the IMOMA which is the improved version of the conventional MOMA. The overall structure of IMOMA is illustrated in Algorithm~\ref{alg:algorithm1}, and corresponding improvement points are as follows. 

\subsubsection{\textbf{Chaotic Solution Initialization}}

\par The chaotic search is a random movement method, which transforms the parameters from the solution space to the chaotic domain to achieve an optimal distribution of initial solutions. The Tent map is utilized to optimize the initial solutions and it is expressed by~\cite{Li2020a}
\begin{equation}
z_{i+1}=\left\{\begin{array}{ll}
z_{i}/{a}, &\quad 0 \leq z_{i} \leq a, \\
(1-z_{i})/({1-a}), &\quad a<z_{i} \leq 1,
\end{array}\right.
\label{con:chaotic}
\end{equation}
\noindent where $z_{i}$ denotes the $i$th value of the Tent chaotic map, and $a\in [0,1]$ denotes the parameter of mapping. Thus, according to the Tent map, the initial solution is calculated by
\begin{equation}
Xr_{i} =lb_r +z_{i} \times \left (ub_r-lb_r\right),
\label{con:initial_r}
\end{equation}
\begin{equation}
Xj_{i} =lb_j +z_{i} \times \left (ub_j-lb_j\right),
\label{con:initial_j}
\end{equation}
\noindent where $Xr_{i}$ and $Xj_{i}$ are the $i$th initial solutions of the relay set and jammer set, which are combined to represent the $i$th solution ($X_{i}$) of the optimization objectives. In addition, $ub_r$ and $lb_r$ are the upper and lower bounds of the relay set, respectively, and $ub_j$ and $lb_j$ are the upper and lower bounds of the jammer set, respectively.

\subsubsection{\textbf{Hybrid Solution Update Strategies}}

\par In this work, inspired by the whale optimization algorithm (WOA), which offers the benefits of simplicity in implementation and high flexibility~\cite{Mirjalili2016}, we introduce a WOA-based solution update strategy to update the positions of UAVs in the jammer set ($Xj$). Then, the arithmetic optimization algorithm (AOA) has the characteristics of fast running speed, low computational complexity and fewer parameters~\cite{Abualigah2021}, which makes it more suitable for optimizing a larger number of UAVs in the relay set. Therefore, we present an AOA-based solution update strategy to update the positions of UAVs in the MUVAA relay ($Xr$). Note that the WOA-based solution update strategy and AOA-based solution update strategy are integrated as the hybrid solution update strategies, which can further balance the exploration and exploitation abilities of the IMOMA. Moreover, for different boundary values in different dimensions, we adopt corresponding update strategies and footsteps. The decision variables in this paper can be updated as follows.

\par \textbf{\textit{First,}} an antenna array can obtain higher gain and avoid mutual coupling when the elements are at appropriate distances in terms of the theories of electromagnetism and CB. Moreover, the UAVs in the MUVAA consume less energy when they are closely distributed in the process of communication~\cite{Sun2022}. Therefore, a better enhancement to the algorithm is to centralize the horizontal positions of UAVs. In this work, we update the horizontal positions of UAVs in the MUVAA jammer ($\mathbb{X}_{j}$, $\mathbb{Y}_{j}$) by using the proposed WOA-based solution update strategy, where whales move either through a shrinking encircling mechanism or a spiral path, with 50\% probability of choosing each~\cite{Mirjalili2016}. The update process is expressed by
\begin{equation}
\begin{split}
\begin{array}{l}
Xj_i^{(\mathbb{X}_{j}, \mathbb{Y}_{j})}=\\
\left\{\begin{array}{ll}
Xj_i^{(\mathbb{X}_{j}, \mathbb{Y}_{j})}+Cj_{i}-l |2r_{2}\cdot Cj_{i}-Xj_i^{(\mathbb{X}_{j}, \mathbb{Y}_{j})}|,&\ p<0.5,\\
Xj_i^{(\mathbb{X}_{j}, \mathbb{Y}_{j})}+| {Cj_{i}-Xj_i^{(\mathbb{X}_{j}, \mathbb{Y}_{j})}}| H +Cj_{i},&\ p\geq 0.5,
\end{array}\right.
\end{array}
\end{split}
\label{j_xy}
\end{equation}
\noindent where $Xj_{i}^{(\mathbb{X}_{j},\mathbb{Y}_{j})}$ denotes the $i$th solution of the jammer set in the horizontal direction, and $Cj_{i}$ = $(\mathbb{E} (\mathbb{X}_{j}), \mathbb{E}(\mathbb{Y}_{j}))$. Moreover, $l$ and $H$ are the related parameters~\cite{Mirjalili2016}, and $r_{2}$ and $p$ are random numbers between 0 and 1. Likewise, inspired by AOA, we utilize a stochastic scaling coefficient to explore diverse regions of the search space and generate more diversification results for the case of more elements. The horizontal positions of UAVs in the MUVAA relay ($\mathbb{X}_{r}$, $\mathbb{Y}_{r}$) can be updated by using the AOA-based solution update strategy, which is as follows:
\begin{equation}
\begin{split}
&Xr_{i}^{(\mathbb{X}_{r},\mathbb{Y}_{r})}=\\
&\left\{\begin{array}{cc}
{Cr}_{i} / ( M^{'}\times((ub_{r}-lb_{r}) \times \mu+lb_{r}) )  , & r_{3}<0.5, \\
{Cr}_{i} \times M\times ((ub_{r}-lb_{r}) \times \mu+lb_{r}), & r_{3}\ge 0.5,
\end{array}\right.
\label{XY_r_multi}
\end{split}
\end{equation}
\noindent where $Xr_{i}^{(\mathbb{X}_{r},\mathbb{Y}_{r})}$ denotes the $i$th solution of the relay set in the horizontal direction, and $Cr_{i}$ = $(\mathbb{E}(\mathbb{X}_{r}),\mathbb{E}(\mathbb{Y}_{r}))$. Moreover, $M$ and $M^{'}$ denote the coefficients of the AOA, $\mu$ denotes the control parameter to tune the search procedure, and $r_{3}$ is the random number between $0$ and $1$. In the AOA, the subtraction and addition as exploitation operators explore the search area deeply on several dense regions to find a better solution. The updating process for the exploitation phase by the AOA-based solution update strategy can be expressed as follows:
\begin{equation}
\begin{split}
&Xr_{i}^{(\mathbb{X}_{r},\mathbb{Y}_{r})}=\\
&\left\{\begin{array}{cc}
Cr_{i} - M\times((ub_{r}-lb_{r}) \times \mu+lb_{r}), &r_{4}<0.5, \\
{Cr}_{i} + M\times((ub_{r}-lb_{r}) \times \mu+lb_{r}), & r_{4}\ge 0.5,
\end{array}\right.
\label{XY_r_add}
\end{split}
\end{equation}
\noindent where $r_{4}$ is the random number between $0$ and $1$.

\par \textbf{\textit{Second,}} as vertical flight costs more energy than horizontal flight, the crucial elements in the third optimization objective are the vertical positions of UAVs in the MUVAA relay (${\mathbb{Z}_{r}}$) and MUVAA jammer (${\mathbb{Z}_{j}}$), which means that the vertical positions of UAVs require being updated more cautiously. Thus, an elite solution of the jammer set ($j$) is chosen from the archive by the roulette wheel selection, and it indicates the fittest UAV in the current optimization to guide the updates of all UAVs. We select $j^{(\mathbb{Z}_{j})}$ to update the solutions of the $z$-axis in the MUVAA jammer by using the WOA-based method, which is as follows:
\begin{equation}
\begin{split}
&Xj_{i}^{(\mathbb{Z}_{j})}=\\
&\left\{\begin{array}{ll}
Xj_i^{(\mathbb{Z}_{j})}+j^{(\mathbb{Z}_{j})}-l |2r_{2}\cdot j^{(\mathbb{Z}_{j})}-Xj_i^{(\mathbb{Z}_{j})}|, &\ p<0.5,\\
Xj_i^{(\mathbb{Z}_{j})}+|{j^{(\mathbb{Z}_{j})}-Xj_i^{(\mathbb{Z}_{j})}}| H +j^{(\mathbb{Z}_{j})}, &\ p \geq 0.5,
\end{array}\right.
\end{split}
\label{j_z}
\end{equation}
\noindent where $Xj_{i}^{(\mathbb{Z}_{j})}$ denotes the $i$th solution of the jammer set in the vertical direction. Likewise, an elite solution of the relay set ($r$) can be chosen from the archive, where $r^{(\mathbb{Z}_{r})}$ is used to update the $z$-axis values in the MUVAA relay by the AOA-based solution update method, which is as follows:
\begin{equation}
\begin{split}
&Xr_{i}^{(\mathbb{Z}_{r})}=\\
&\left\{\begin{array}{cc}
r^{(\mathbb{Z}_{r})} / ( M^{'}\times( (ub_{r}-lb_{r})\times \mu+lb_{r})), &r_{3}<0.5,\\
r^{(\mathbb{Z}_{r})} \times M\times ((ub_{r}-lb_{r})\times \mu+lb_{r}), &r_{3}\ge 0.5,
\end{array}\right.
\label{Z_r_multi}
\end{split}
\end{equation}
\begin{equation}
\begin{split}
&Xr_{i}^{(\mathbb{Z}_{r})}=\\
&\left\{\begin{array}{cc}
r^{(\mathbb{Z}_{r})} - M\times((ub_{r}-lb_{r}) \times \mu+lb_{r}), &r_{4}<0.5,\\
r^{(\mathbb{Z}_{r})} + M\times((ub_{r}-lb_{r}) \times \mu+lb_{r}), &r_{4}\ge 0.5,
\end{array}\right.
\label{Z_r_add}
\end{split}
\end{equation}
\noindent where $Xr_{i}^{(\mathbb{Z}_{r})}$ denotes the $i$th solution of the relay set in the vertical direction after the update.

\par \textbf{\textit{Finally,}} proper excitation current weights in the MUVAA relay ($\mathbb{I}_{r}$) can efficiently adjust the beam pattern when the UAVs in the MUVAA relay communicate with the legitimate vessel. Moreover, proper excitation current weights in the MUVAA jammer ($\mathbb{I}_{j}$) can assist UAVs in delivering stronger jamming signals to the eavesdropper, thus enhancing communication rates and improving security performance. Therefore, we choose the elite solution of the jammer set $j^{(\mathbb{X}_{j},\mathbb{Y}_{j},\mathbb{Z}_{j})}$ to substitute the previous corresponding solutions $Xj_{i}^{(\mathbb{X}_{j},\mathbb{Y}_{j},\mathbb{Z}_{j})}$, and use $j^{(\mathbb{I}_{j})}$ to iterate over $Xj_{i}^{(\mathbb{I}_{j})}$. The $i$th solution in the MUVAA jammer $Xj_{i}^{(\mathbb{I}_{j})}$ by using the WOA-based method can be updated as follows:
\begin{equation}
\begin{split}
&Xj_i^{(\mathbb{I}_{j})}=\\
&\left \{\begin{array}{ll}
Xj_i^{(\mathbb{I}_{j})}+j^{(\mathbb{I}_{j})}-l|2r_{2}\cdot j^{(\mathbb{I}_{j})}-j^{(\mathbb{I}_{j})}|, & p<0.5.\\
Xj_i^{(\mathbb{I}_{j})}+|{j^{(\mathbb{I}_{j})}-Xj_i^{(\mathbb{I}_{j})}}|H+j^{(\mathbb{I}_{j})}, & p\geq0.5.
\end{array}\right.
\end{split}
\label{j_I}
\end{equation}
\noindent Moreover, the elite solution of the relay set $r^{(\mathbb{X}_{r},\mathbb{Y}_{r},\mathbb{Z}_{r})}$ can be employed to substitute the previous corresponding solutions $Xr_{i}^{(\mathbb{X}_{r},\mathbb{Y}_{r},\mathbb{Z}_{r})}$. The $i$th solution in the MUVAA relay $Xr_{i}^{(\mathbb{I}_{r})}$ can be iterated with $r^{(\mathbb{I}_{r})}$ by using the AOA-based method as follows:
\begin{equation}
\begin{split}
&Xr_{i}^{(\mathbb{I}_{r})}=\\
&\left\{\begin{array}{cc}
r^{(\mathbb{I}_{r})} / ( M^{'}\times((ub_{r}-lb_{r}) \times \mu+lb_{r})), & r_{3}<0.5.\\
r^{(\mathbb{I}_{r})} \times M\times((ub_{r}-lb_{r}) \times \mu+lb_{r}), & r_{3}\ge 0.5.
\end{array}\right.
\label{I_r_multi}
\end{split}
\end{equation}
\begin{equation}
\begin{split}
&Xr_{i}^{(\mathbb{I}_{r})}=\\
&\left\{\begin{array}{cc}
r^{(\mathbb{I}_{r})}- M\times((ub_{r}-lb_{r})\times \mu+lb_{r}), & r_{4}<0.5. \\
r^{(\mathbb{I}_{r})} + M\times((ub_{r}-lb_{r}) \times \mu+lb_{r}), & r_{4}\ge 0.5.
\end{array}\right.
\label{I_r_add}
\end{split}
\end{equation}

\begin{algorithm}
\label{alg:algorithm1}
\caption{IMOMA}
\KwIn {Population size ${N}$, maximum iteration ${t}_{max}$, archive set $Ar$; $\#$ \boldsymbol{$Ar$} \textbf{for storing the Pareto solutions.}}
Set the corresponding parameters; \\
\For{$i=1$ to ${N}$}{
Initialize the $i$th solution of the relay set ($Xr_{i}$) by Eq. (\ref{con:initial_r});\\
Initialize the $i$th solution of the jammer set ($Xj_{i}$) by Eq. (\ref{con:initial_j});\\
}
\For{$t=1$ to ${t}_{max}$} {
Compute the optimization objective values and update $Ar$ based on non-dominated solutions;\\
\For{ $i=1$ to ${N}$}{
Calculate the $Xr_{i}$ and $Xj_{i}$ of the $i$th mayfly relay set and jammer set (Note that $Xr_{i}$ and $Xj_{i}$ are integrated as $X_{i}$) by Eq. (\ref{con:MOMA});\\
Compute $\zeta$ by Eq. (\ref{con:zeta}); \textbf{$\#$ Threshold \boldsymbol{$\zeta$} for updating the relay and jamming sets.}\\
Generate some random numbers ($r_{1}$, $r_{3}$ and $r_{4}$) between 0 and 1;\\
Update the $Xj_{i}$ by using \textbf{Algorithm~\ref{alg:algorithm2}};\\
Update the $Xr_{i}$ by using \textbf{Algorithm~\ref{alg:algorithm3}};\\
}
}	
\KwOut {Updated $Ar$.}
\end{algorithm}

\par Accordingly, the IMOMA is shown in Algorithm \ref{alg:algorithm1}, in which the marker $\zeta$ is a threshold used for regulating and it is calculated by~\cite{Sun2021a}:
\begin{equation}
\zeta=\left\{\begin{array}{lc}
0.5-\frac{t}{t_{\max }}, & t<\frac{t_{\max}}{2}. \\
\frac{t}{t_{\max }}-0.5, & \text {otherwise}.
\end{array}\right.
\label{con:zeta}
\end{equation}

\begin{algorithm}
\label{alg:algorithm2}
\caption{WOA-based solution update algorithm of the jammer set}
\KwIn { Current jammer set $Xj_{i}$, current elite solution of jammer set $j$; \textbf{$\#$ Elite solution is the current most appropriate solution.}}
Update the value of $Xj_{i}^{(\mathbb{Z}_{j})}$ by Eq. (\ref{j_z});\\
\eIf{$t<t_{max}/{2}$}
{
\If{$r_{1}< \zeta$}{
Update the value of $Xj_{i}^{(\mathbb{X}_{j},\mathbb{Y}_{j})}$ by Eq. (\ref{j_xy});\\
}
}
{
\If{$r_{1}< \zeta $}{
Substitute $Xj_{i}^{(\mathbb{X}_{j},\mathbb{Y}_{j},\mathbb{Z}_{j})}$ with the value of $j^{(\mathbb{X}_{j},\mathbb{Y}_{j},\mathbb{Z}_{j})}$;\\
Update the value of $Xj_{i}^{(\mathbb{I}_{j})}$ by Eq. (\ref{j_I});\\
}
}
\KwOut { Updated jammer set $Xj_{i}^{(\mathbb{X}_{j},\mathbb{Y}_{j},\mathbb{Z}_{j},\mathbb{I}_{j})}$;}
\end{algorithm}

\par Moreover, the AOA-based solution update algorithm of the relay set is shown in Algorithm \ref{alg:algorithm3}. The math optimizer accelerated (MOA) function can be used to select the search phase (i.e., exploration or exploitation), which is calculated by~\cite{Zervoudakis2020}:
\begin{equation}
    MOA(t)=Min+t\times (\frac{Max-Min}{t_{max}} ),
\label{MOA}
\end{equation}
\noindent where $MOA(t)$ denotes the function value at the $t$th iteration. Moreover, $t$ is the current iteration, ranging from 1 to $t_{max}$. In addition, $Min$ and $Max$ denote the minimum and maximum values of the accelerated function, respectively.

\begin{algorithm}
\label{alg:algorithm3}
\caption{AOA-based solution update algorithm of the relay set}
Compute the value of $MOA$ by Eq. (\ref{MOA});\\
\textbf{Input:} Current relay set $Xr_{i}$, current elite solution of relay set $r$;\\
\eIf{$r_{1}>MOA$}{
Exploration phase: Determine applying multiplication or division  operator, update $Xr_{i}^{(\mathbb{Z}_{r})}$ by Eq. (\ref{Z_r_multi});\\
}
{
Exploitation phase: Determine applying addition or subtraction  operator, update $Xr_{i}^{(\mathbb{Z}_{r})}$ by Eq. (\ref{Z_r_add});\\
}
\eIf{$t<t_{max}/{2}$}
{
\If{$r_{1}< \zeta$}{
    \eIf{$r_{1}>MOA$}{Update $Xr_{i}^{(\mathbb{X}_{r},\mathbb{Y}_{r})}$ by Eq. (\ref{XY_r_multi}); $\#$\textbf{Exploration phase.}\\
    }
    {
Update $Xr_{i}^{(\mathbb{X}_{r},\mathbb{Y}_{r})}$ by Eq. (\ref{XY_r_add}); $\#$\textbf{Exploration phase.}\\
    }
}
}
{
\If{$r_{1}< \zeta $}{
Substitute $Xr_{i}^{(\mathbb{X}_{r},\mathbb{Y}_{r},\mathbb{Z}_{r})}$ with $r^{(\mathbb{X}_{r},\mathbb{Y}_{r},\mathbb{Z}_{r})}$;\\
\eIf{$r_{1}>MOA$}{
Update $Xr_{i}^{(\mathbb{I}_{r})}$ by Eq. (\ref{I_r_multi}); $\#$\textbf{Exploration phase.}\\
}
{
Update $Xr_{i}^{(\mathbb{I}_{r})}$ by Eq. (\ref{I_r_add}); $\#$\textbf{Exploration phase.}\\
}
}
}
\KwOut {Updated relay set $Xr_{i}^{(\mathbb{X}_{r},\mathbb{Y}_{r},\mathbb{Z}_{r},\mathbb{I}_{r})}$;}
\end{algorithm}

\subsection{Complexity of the IMOMA}

\par The computational complexity of the proposed algorithm mainly depends on the computations of the optimization objectives and sorting the solutions in each optimization objective. We denote the number of optimization objectives, population size, and archive size as $N_{obj}$, $N$, and $N_{a}$, respectively. Specifically, the optimization objective computation has $ \mathcal{O}(N_{obj}\cdot N)$ computational complexity. Moreover, for sorting the solutions in each objective, the computational complexity of classifying the $N_{a}$ solutions in the Pareto archive is $ \mathcal{O} (N_{obj}\cdot N_{a}\cdot \log N_{a})$. In this paper, we set $N_{a}$ to the same size as $N$, then the computational complexity for the non-dominated sorting is $ \mathcal{O} (N_{obj}\cdot N^{2})$, and the overall complexity of the proposed IMOMA is $\mathcal{O} (N_{obj}\cdot N^{2})$.

%

\section{Simulation results and analysis} \label{sec:simulation results and analysis}

\par In this section, the performance of the proposed improved algorithm is evaluated by the simulations. 

\subsection{Simulation Setups}

\subsubsection{Parameter Settings}

\par The simulation experiments are conducted using Matlab 9.2. The 3D positions (in meters) of Bob and Willie, which are set to (2400, 2300, 5) and (2000, 2000, 5), respectively, and the sea level is set as 5 m. The distribution area of the UAV relay set ($A_{sr}$) and UAV jammer set ($A_{sj}$) are located within a 100 m $\times$ 100 m area to form the MUVAA relay and MUVAA jammer. We randomly initialize the hovering positions of the UAVs from their feasible flight area since they may have been working on other tasks before. Moreover, we consider a larger scale network including 16 and 8 UAVs in the relay set and jammer set, and a smaller scale network with 8 and 4 UAVs in the relay set and jammer set. In addition, the remaining key parameters used in the simulations are shown in Table \ref{table:table1}~\cite{Wang2021}~\cite{Huang2024}.

\subsubsection{Baselines} To demonstrate the effectiveness of the proposed IMOMA, three comparison approaches and various comparison algorithms are introduced as follows.

\begin{itemize}
    \item \textit{Non-CB Approach:} This approach does not use CB to achieve signal transmission. Specifically, a UAV, denoted as UAV-R, acts as a relay to forward data signals from the LBS to Bob by the data link, the other UAV, denoted as UAV-J, moves from its hovering position towards Willie and sends jamming signals by the jamming link at a suitable location. As such, the comparison approach can highlight the effect of CB in long-distance signal transmission.
    
    \item \textit{Single CB Approach:} This approach only utilizes CB to send data signals from the MUVAA relay to Bob, and UAV-J sends jamming signals to Willie. In this case, the comparison between this approach and the proposed method can illustrate the effectiveness of CB in long-distance data signal transmission and the necessity of establishing an MUVAA jammer.
    
    \item \textit{Multi-hop Approach:} The approach employs UAV multi-hop to achieve data and jamming signal transmission, which is used to extend communication range. As such, the comparison between this approach and the proposed method further highlights the effectiveness of CB in long-distance transmission and its capability for efficient energy savings.
    
    \item \textit{State-of-the-art Swarm Intelligence Algorithms:} We select conventional MOMA, multi-objective dragonfly algorithm (MODA)~\cite{Mirjalili2016a}, multi-objective multi-verse optimization (MOMVO)~\cite{Mirjalili2017a} and multi-objective ant lion optimizer (MALO)~\cite{Mirjalili2017} as benchmark swarm intelligence algorithms for comparison. These algorithms are known for their excellent diversity and exploration capabilities in solving MOP, and their efficiency has been well-validated by existing studies~\cite{Li2024}~\cite{Sun2021a}. Specifically, MODA simulates the social behavior of dragonflies to preserve  diversity, MOMVO extends the search space based on the parallel universe theory, and MALO strikes a balance between exploration and exploitation by mimicking the predatory behavior of ant lions. The rich variety of mechanisms provides a robust set of benchmarks that highlights the effectiveness of the proposed algorithm in addressing the formulated problem. The maximum number of iterations and population size in these aforementioned algorithms are set as 500 and 30, respectively.

\end{itemize}

\par Furthermore, we provide the comparison results of different baselines and the CB-based approach.

\begin{table}[!t]
\renewcommand{\arraystretch}{1.3}
\newcommand{\tabincell}[2]{\begin{tabular}{@{}#1@{}}#2\end{tabular}}
\caption{Summary of main notations}
\label{table:notations}
\centering
\renewcommand\arraystretch{1.3}
\begin{tabular}{cc}
\hline
\bfseries Notation & \bfseries Meaning \\
\hline
 & Notation in system model\\
\hline
$\theta$ & Elevation angle between the MUVAA relay and vessel\\

$\phi$ & Azimuth angle between the MUVAA relay and vessel\\

$AF_{r}(\cdot)$  & Array factor of the MUVAA relay\\

$\lambda$ & Wavelength\\

$G_{v}(\cdot) $& Antenna gain of the MUVAA relay\\

$PL(\cdot)$ & Path loss from the MUVAA relay\\

$d$ & Distance from the center of MUVAA relay to vessel\\

$\theta^{'}$ & Elevation angle between the MUVAA jammer and vessel\\

$\phi^{'}$ & Azimuth angle between the MUVAA jammer and vessel\\

$AF_{j}(\cdot)$  & Array factor of the MUVAA jammer\\

$G_{v}^{'}(\cdot)$ & Antenna gain of the MUVAA jammer\\

$PL^{'}(\cdot)$ & Path loss from the MUVAA jammer\\

$d^{'}$ & Distance from the center of MUVAA jammer to vessel\\

$v_{m}$ & Mean rotor induced velocity in hovering\\

$v_{t}$  & Tip speed of the rotor blade\\

$d_{f}$ & Fuselage drag ratio\\

$s_{r}$ & Rotor solidity\\

$\rho_{a}$ & Air density \\

$a_{r}$ & Rotor disc area \\
\hline
 & Notation in the IMOMA\\
\hline
$N$ & Population size\\ 

$ub_{r}$ & Upper bound of relay set\\

$lb_{r}$ & Lower bound of relay set \\

$ub_{j}$ & Upper bound of jammer set \\

$lb_{j}$ & Lower bound of jammer set  \\

$M$ & Coefficients of the AOA \\ 

$M^{'}$ & Coefficients of the AOA \\ 

$\zeta $ &  Threshold \\ 

$a$ & Parameter of Tent mapping \\

$\mu$ & Control parameter of the AOA \\

$j$ & An elite solution of jammer set \\ 

$r$ & An elite solution of relay set\\ 

$Max$ & Maximum value of the accelerated function of the AOA \\

$Min$ & Minimum value of the accelerated function of the AOA \\
\hline
\end{tabular}
\end{table}

\begin{table}[!t]
\renewcommand{\arraystretch}{1.3}
\newcommand{\tabincell}[2]{\begin{tabular}{@{}#1@{}}#2\end{tabular}}
\caption{Main parameters in the simulation process}
\label{table:table1}
\centering
\renewcommand\arraystretch{1.3}
\begin{tabular}{ccc}
\hline
\bfseries Notation & \bfseries Meaning & \bfseries Default value\\
\hline
$f_{c}$ & Carrier frequency & 2.4 GHz\\
\hline
$P_{UR}$ & \tabincell{c} {Transmission power of each UAV \\ in the MUVAA relay} & 0.1 W\\
\hline
$P_{UJ}$ & \tabincell{c} {Transmission power of each UAV \\ in the MUVAA jammer} & 0.1 W\\
\hline
$\sigma^{2}$ & Power of additive white Gaussian noise & -150 dBm\\
\hline
$m_{U}$ & Aircraft mass & 2 kg\\
\hline	
$\eta _{LOS}$ & Attenuation factor for LoS links & 2.3 dB\\
\hline
$\eta _{NLOS}$ & Attenuation factor for NLoS links & 34 dB\\
\hline
$\alpha_{a}$ & Sigmoid function parameter & 5.0188 dB\\\hline
$\alpha_{b}$ & Sigmoid function parameter & 0.3511 dB\\
\hline
$C_U$ & Constant parameter & 1 dB \\
\hline
$C_{r}$ & Environment-related parameter & 34 dB\\
\hline
$Lr_{min}$ & Minimum horizontal scope of $\mathcal{U}_{R}$ & 0 m\\
\hline
$Lr_{max}$ & Maximum horizontal scope of $\mathcal{U}_{R}$ & 100 m\\
\hline
$Lj_{xmin}$ & Minimum x-axis horizontal scope of $\mathcal{U}_{J}$ & 4400 m\\
\hline
$Lj_{xmax}$ & Maximum x-axis horizontal scope of $\mathcal{U}_{J}$ & 4500 m\\
\hline
$Lj_{ymin}$ & Minimum y-axis horizontal scope of $\mathcal{U}_{J}$ & 4300 m\\
\hline
$Lj_{ymax}$ & Maximum y-axis horizontal scope of $\mathcal{U}_{J}$ & 4400 m\\
\hline
$H_{min}$ & Minimum vertical scope of $\mathcal{U}_{R}$ and $\mathcal{U}_{J}$ & 60 m\\
\hline
$H_{max}$ & Maximum vertical scope of $\mathcal{U}_{R}$ and $\mathcal{U}_{J}$ & 120 m\\
\hline
$P_{R}$ & Transmission power of the UAV-R & 0.1 W\\
\hline
$P_{J}$ & Transmission power of the UAV-J & 0.1 W\\
\hline
\end{tabular}
\end{table}

\begin{figure}
    \centering
   \subfigure[Gain distributions of the MUVAA relay.]{
       \includegraphics[width=0.48\linewidth]{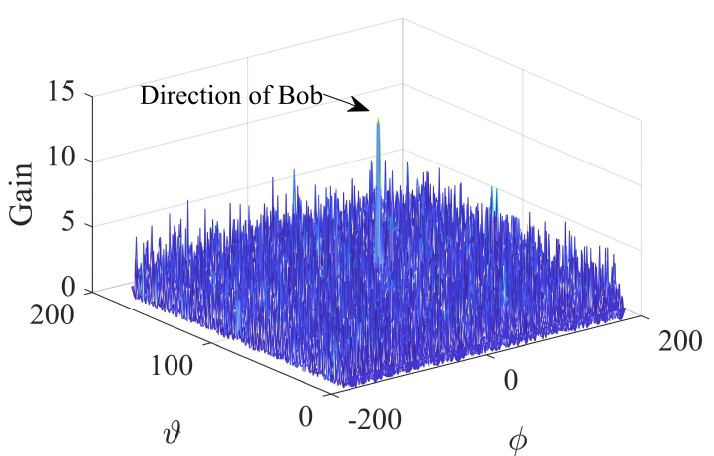}}\hfill
   \subfigure[Gain distributions of the MUVAA jammer.]{
        \includegraphics[width=0.48\linewidth]{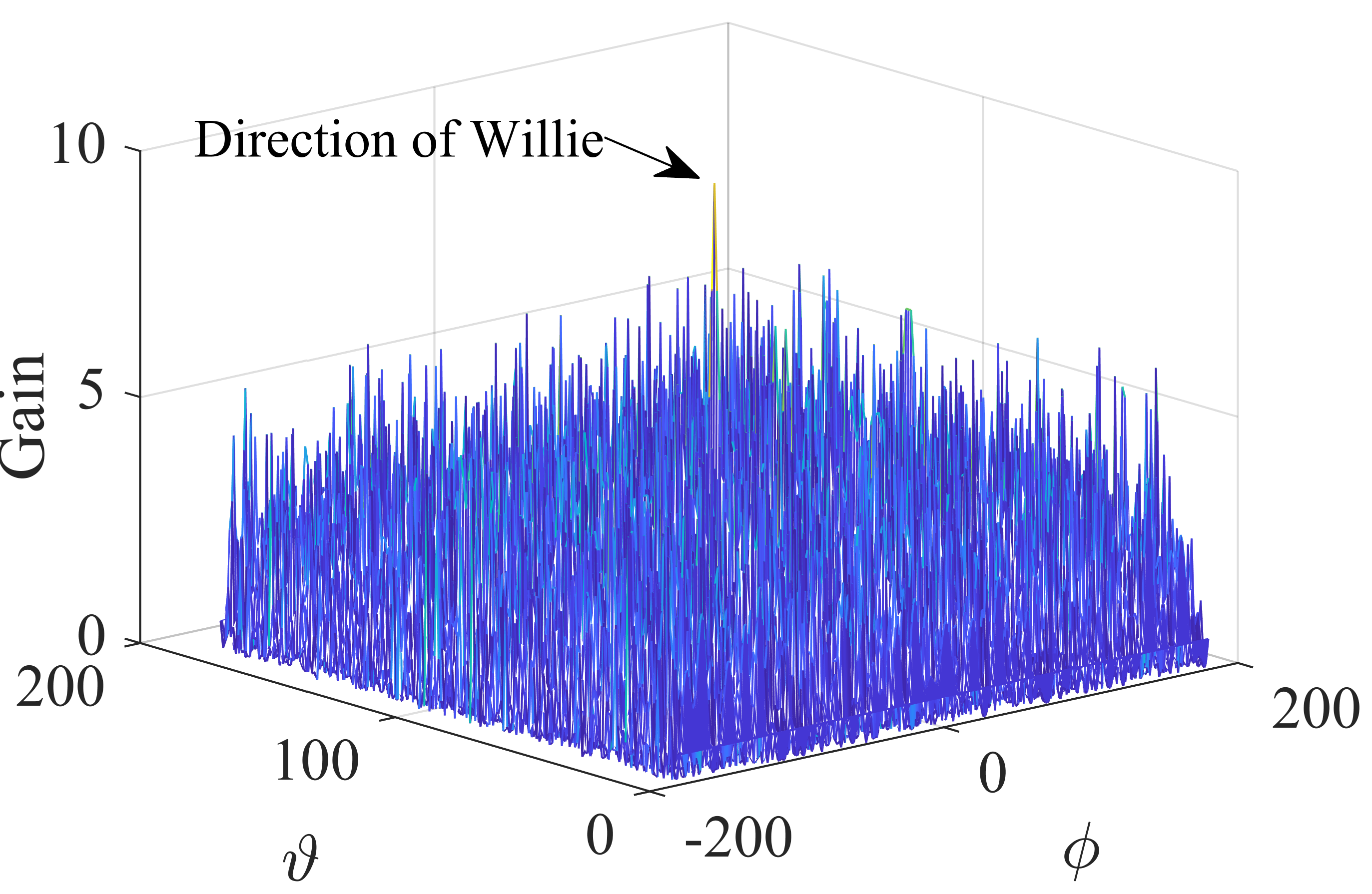}}\hfill
   \caption{Gain distributions optimized by the IMOMA in larger scale network.}
 \label{results1}
\end{figure}

\begin{figure*}
	\centering
	\subfigure[Movement paths of UAVs in the MUVAA relay.]{\includegraphics[scale=0.43]{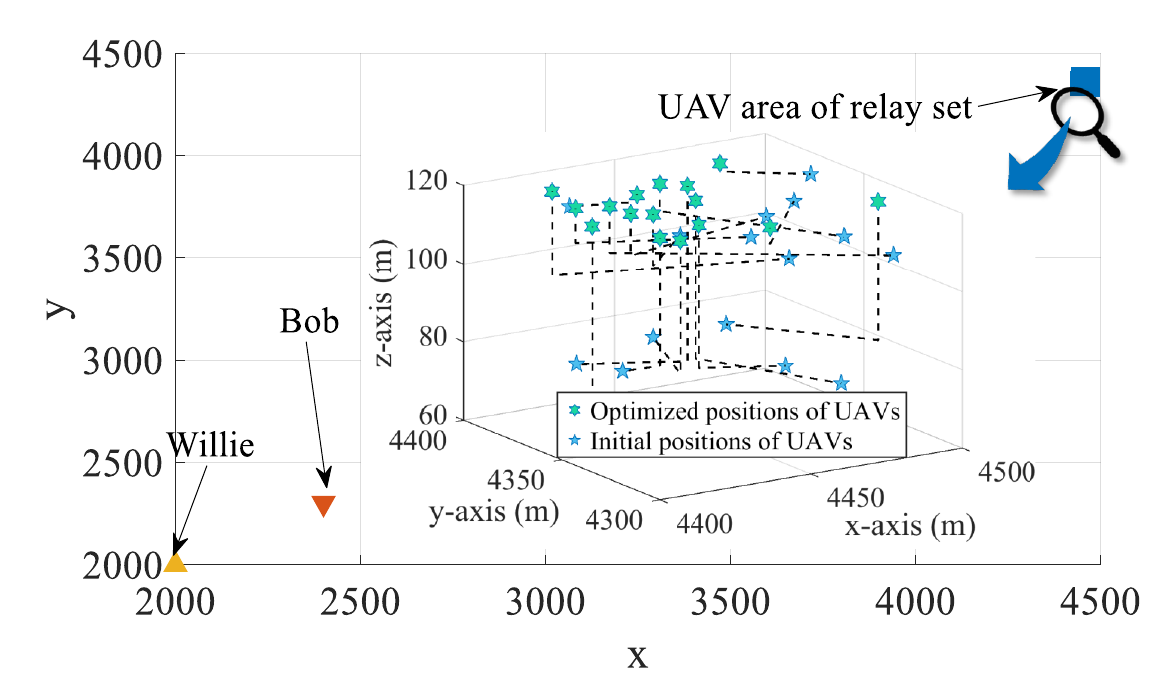}}\hfill
	\subfigure[Movement paths of UAVs in the MUVAA jammer.]{\includegraphics[scale=0.43]{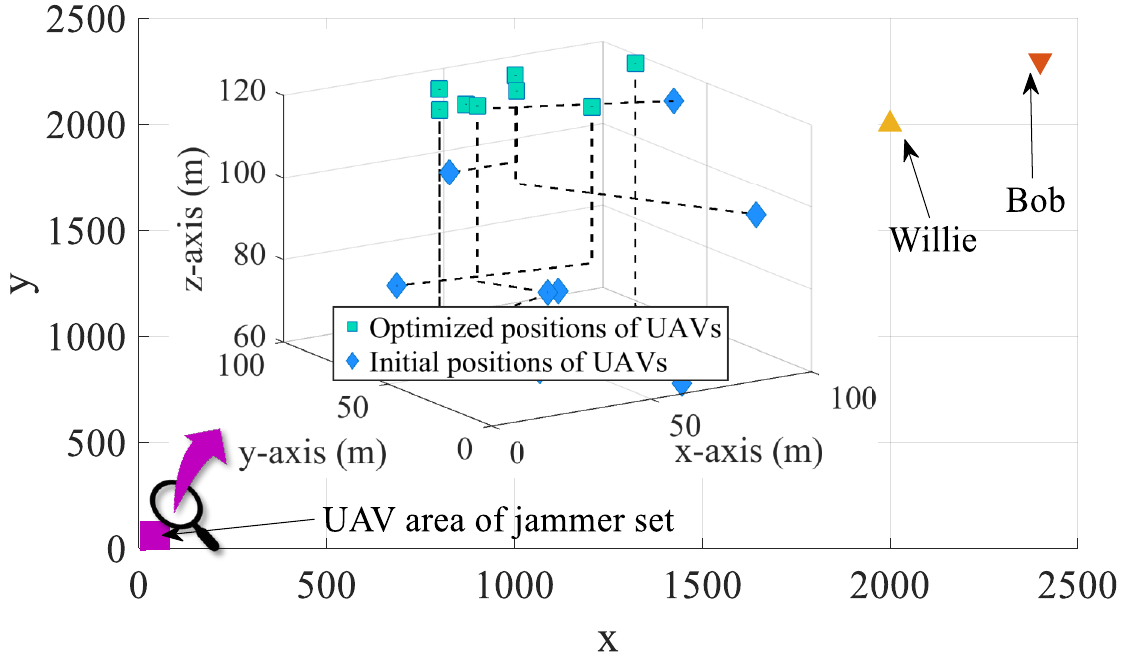}}
	\caption{Movement paths optimized by the IMOMA in larger scale network.}
 \label{results2}
\end{figure*}

\subsection{Simulation Results}

\subsubsection{Visualization Results} 

\par This part presents the visualization results of the larger scale network.

\par Fig. \ref{results1} shows the distribution of antenna gains optimized by the proposed IMOMA. Specifically, Fig. \ref{results1}(a) shows the antenna gains from the MUVAA relay in all directions. As can be seen, the gain towards Bob is the highest among all directions, which makes Bob receives the maximum strength of data signals. Fig. \ref{results1}(b) shows the antenna gain from the MUVAA jammer in all directions. It can be seen that Willie receives the maximum strength of jamming signals. Thus, the legitimate vessel Bob can achieve more secure and reliable maritime communication performance. Moreover, Fig. \ref{results2} illustrates the movement paths of UAVs in the MUVAA relay and MUVAA jammer from the initial hovering positions to the optimized positions, which are obtained by the proposed IMOMA. As can be seen, in the MUVAA relay and jammer, the optimized UAV positions are more centralized and compact than the original positions, which results in stronger transmitted signals and optimal SINR values compared to less centralized positions of UAVs. This more focused placement facilitates better implementation of CB, thus achieving more energy-efficient maritime wireless communications.

\begin{figure}[!t]
\centering
\includegraphics[width=3.5in]{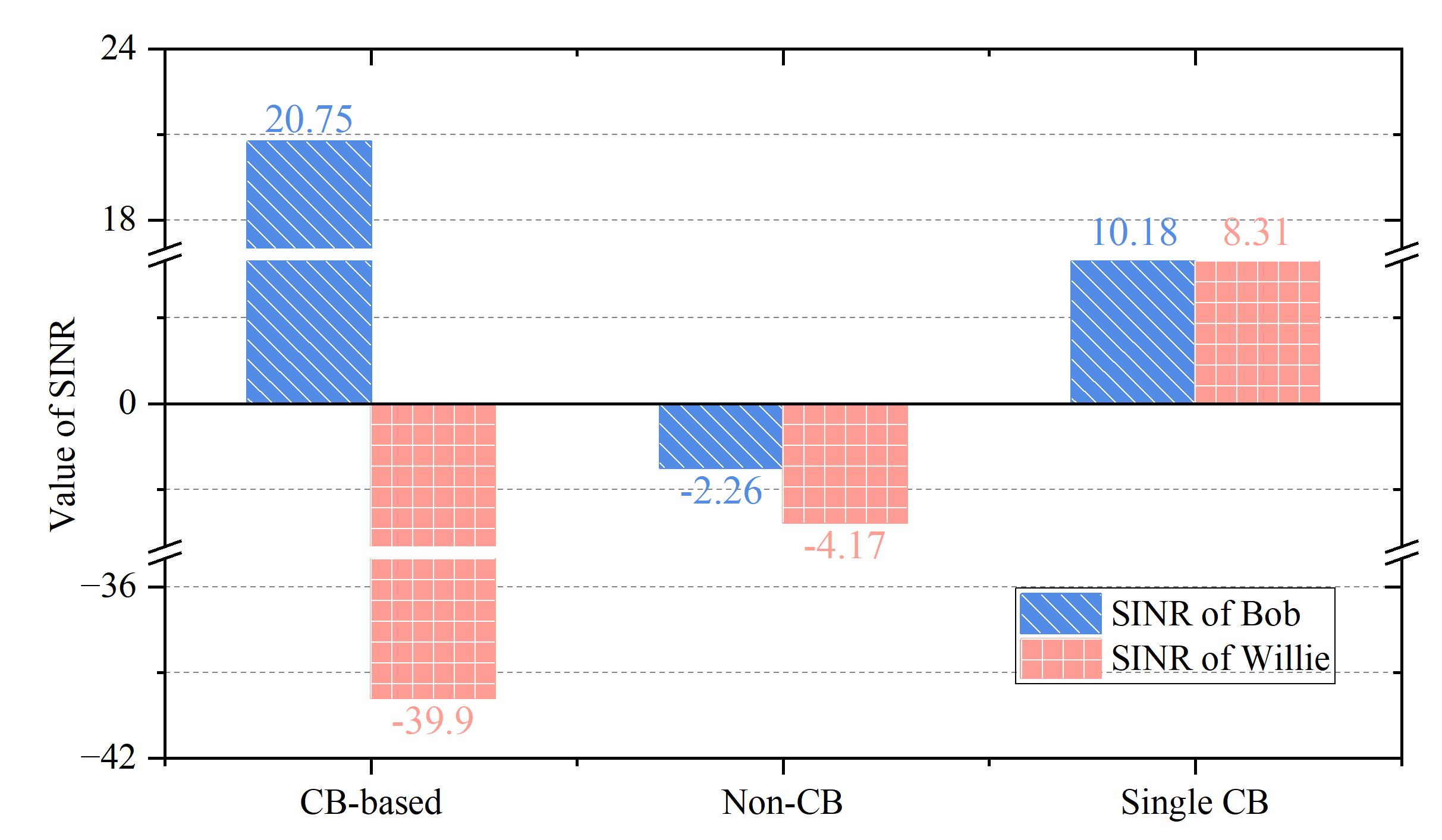}
\caption{The values of SINR of Bob and Willie obtained by the approaches of CB-based, non-CB, and single CB.}
\label{value_compare12}
\end{figure}

\begin{figure}[!t]
\centering
\includegraphics[width=3.5in]{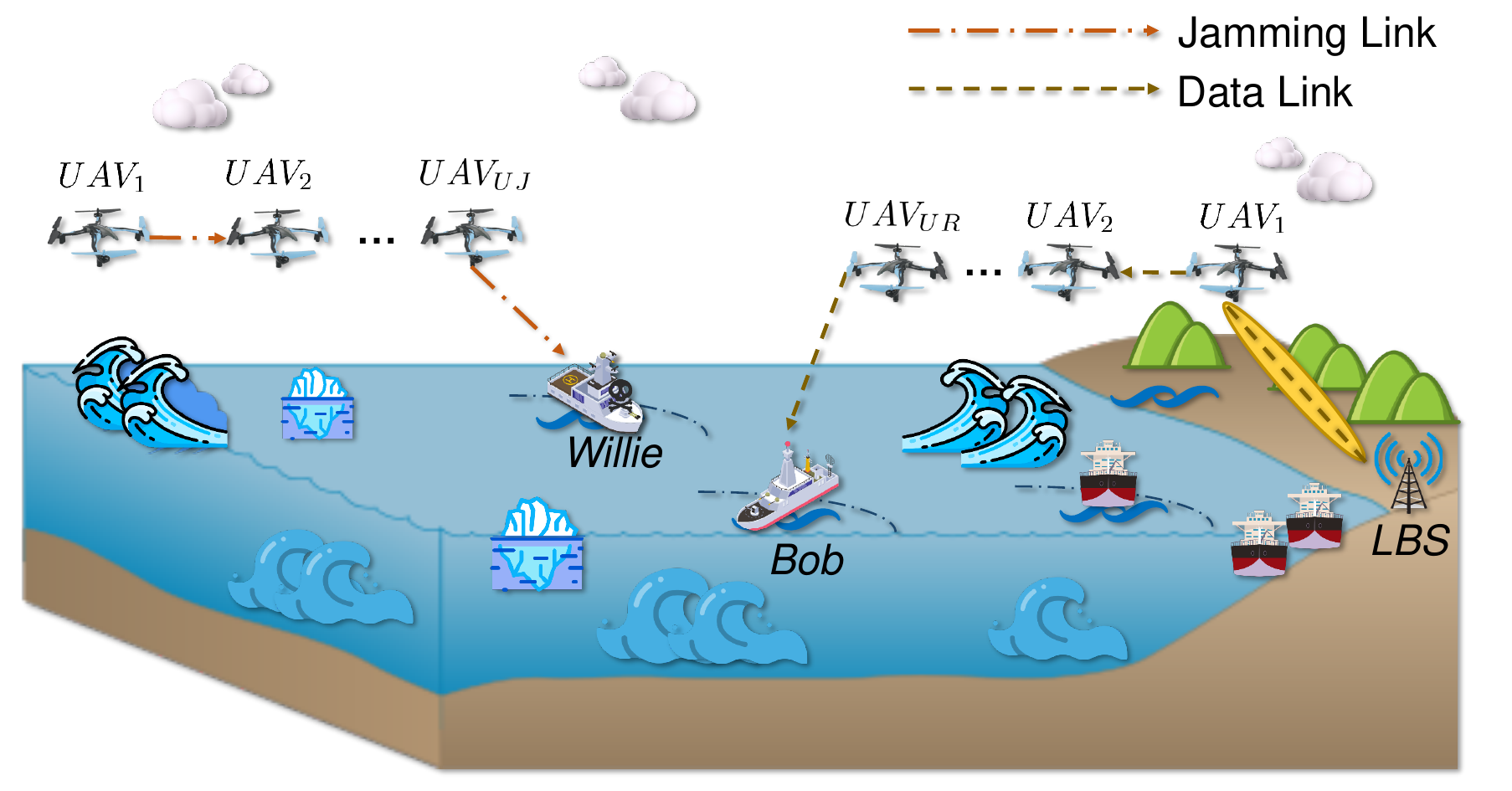}
\caption{A UAV multi-hop maritime communication system.}
\label{multi-hop}
\end{figure}

\begin{table}[!t]
\renewcommand{\arraystretch}{1.3}
\newcommand{\tabincell}[2]{\begin{tabular}{@{}#1@{}}#2\end{tabular}}
\caption{Performance comparison between CB-based and multi-hop methods}
\label{table:multi-hop}
\centering
\renewcommand\arraystretch{1.3}
\begin{tabular}{c|ccc|ccc}
\hline
\bfseries \multirow{2}{*}{Methods}   &\multicolumn{3}{c|}{ Smaller scale network} \bfseries  & \multicolumn{3}{c}{Larger scale network}  \\
\cline{2-7} 
& $f_{1}$ &$f_{2}$ & $f_{3}$ (J) & $f_{1}$ & $f_{2}$ & $f_{3}$ (J)\\
\hline
Multi-hop & 0 &0 &4.1$\times 10^{5}$  & 0 & 0 &5.2$\times 10^{6}$\\
\textbf{CB-based} &\textbf{15.5} & \textbf{-27.9} &\pmb{6.6$\times 10^{4}$} & \textbf{20.8} &\textbf{-39.9} &\pmb{1.4$\times 10^{5}$} \\
\hline
\end{tabular}
\end{table}

\subsubsection{Comparison with the Different Approaches} 

\par Fig. \ref{value_compare12} shows the optimization objective values in terms of $f_{1}$ and $f_{2}$, which are obtained by the CB-based, non-CB, and single CB approaches. First, the satisfactory values SINR of Bob and Willie under the CB-based approach in larger scale network indicate that CB can achieve remote maritime transmission and effectively protect against eavesdropping. Second, the SINR of Bob of the non-CB approach is negative, which suggests that Bob cannot effectively receive information by the data link, and the non-CB approach can not achieve long-distance communications. Moreover, data signals are unlikely to be received by Willie next to Bob. Finally, according to the single CB approach, Bob and Willie both can receive data signals, which demonstrates that data signals can be sent by CB, and the jamming signals of non-CB can not affect Willie and can not guarantee the safe transmission of data signals.

\par Furthermore, we use the UAV multi-hop method to compare with the CB-based approach. Fig. \ref{multi-hop} shows the sketch of the UAV multi-hop communication system. Specifically, multiple UAVs are uniformly deployed between the initial area and the target location at the same altitude~\cite{Chen2018}. The free-space channel model is applied to represent airborne communications, and the results are shown in Table \ref{table:multi-hop}. As can be seen, the SINR values of Bob and Willie obtained by the multi-hop method are 0, which is attributed to the long-distance transmission preventing the target user from receiving the signals. These comparison results further demonstrate the effectiveness of the CB approach. In summary, the results mentioned above indicate that the CB-based approach outperforms other approaches, achieving more efficient maritime secure communications. In addition, as shown in Fig. \ref{results1}, the CB-based approach can maximize the antenna gain in the target direction, which further illustrates its reliability and efficiency.

\par In this paper, the implementation of the key CB approach requires increasing the number of UAVs. In this case, the VAA formed by multiple UAVs is used to improve signal coverage and transmission reliability, whereas the process inevitably leads to more energy consumption. Note that this trade-off is necessary to achieve more efficient maritime secure communications. In real-time communication requirements, communication effectiveness is the main concern, and the CB method can substantially improve the long-distance signal transmission quality and ensure security. Therefore, despite the rise in energy consumption, the CB approach is reasonable and in line with the realistic needs.

\subsubsection{Comparison with other Algorithms}

\par
\begin{figure}
    \centering
   \subfigure[Solution distributions in larger scale network.]{
       \includegraphics[width=0.9\linewidth]{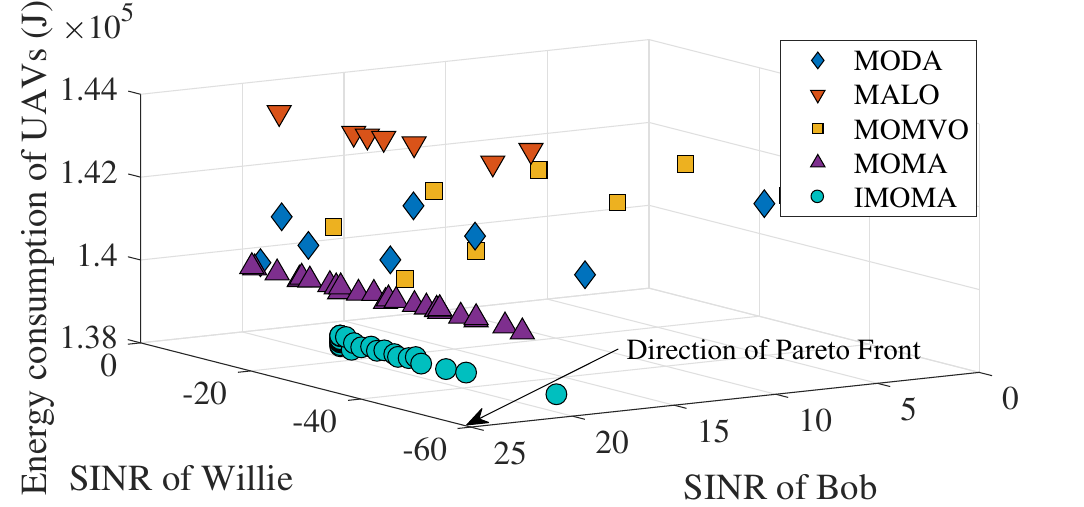}}
    \label{fig:cost_large}\\
   \subfigure[Solution distributions in smaller scale network.]{
        \includegraphics[width=0.9\linewidth]{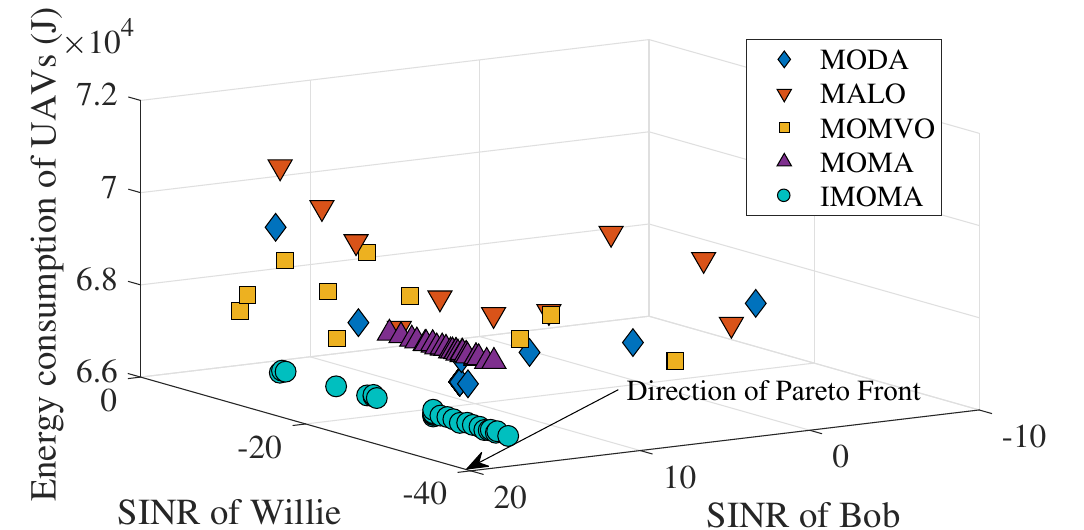}}
    \label{fig:cost_small}\\
   \caption{Solution distributions obtained by different algorithms in larger and smaller scale networks of CB-based approach.}
   \label{fig:solutions distributions}
\end{figure}

 Fig. \ref{fig:solutions distributions} shows the Pareto solution distributions obtained by the different algorithms in larger and smaller scale networks of the CB-based approach. The coordinates of the points in the three-dimensional space represent the values derived from the three optimization objectives. It is obvious that the solutions obtained by the proposed IMOMA are more concentrated and they are closer to the PF both in larger and smaller scale networks. Therefore, the proposed IMOMA is more suitable for forming VAA and has greater superiority in solving the corresponding optimization problems.

\par
\begin{figure}[!t]
\centering
 \subfigure[The optimization objective values in larger scale network.]{
\includegraphics[width=3.5in]{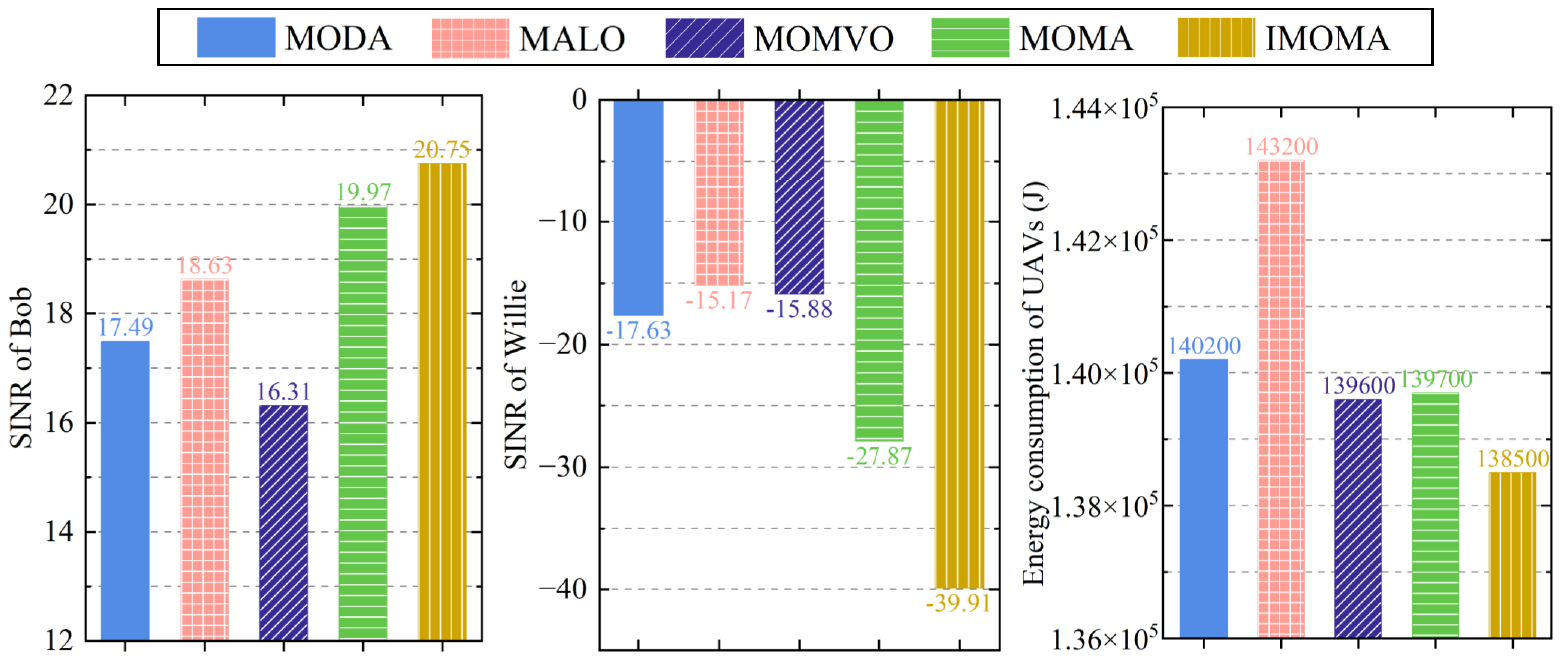}}\\
 \subfigure[The optimization objective values in smaller scale network.]{
\includegraphics[width=3.5in]{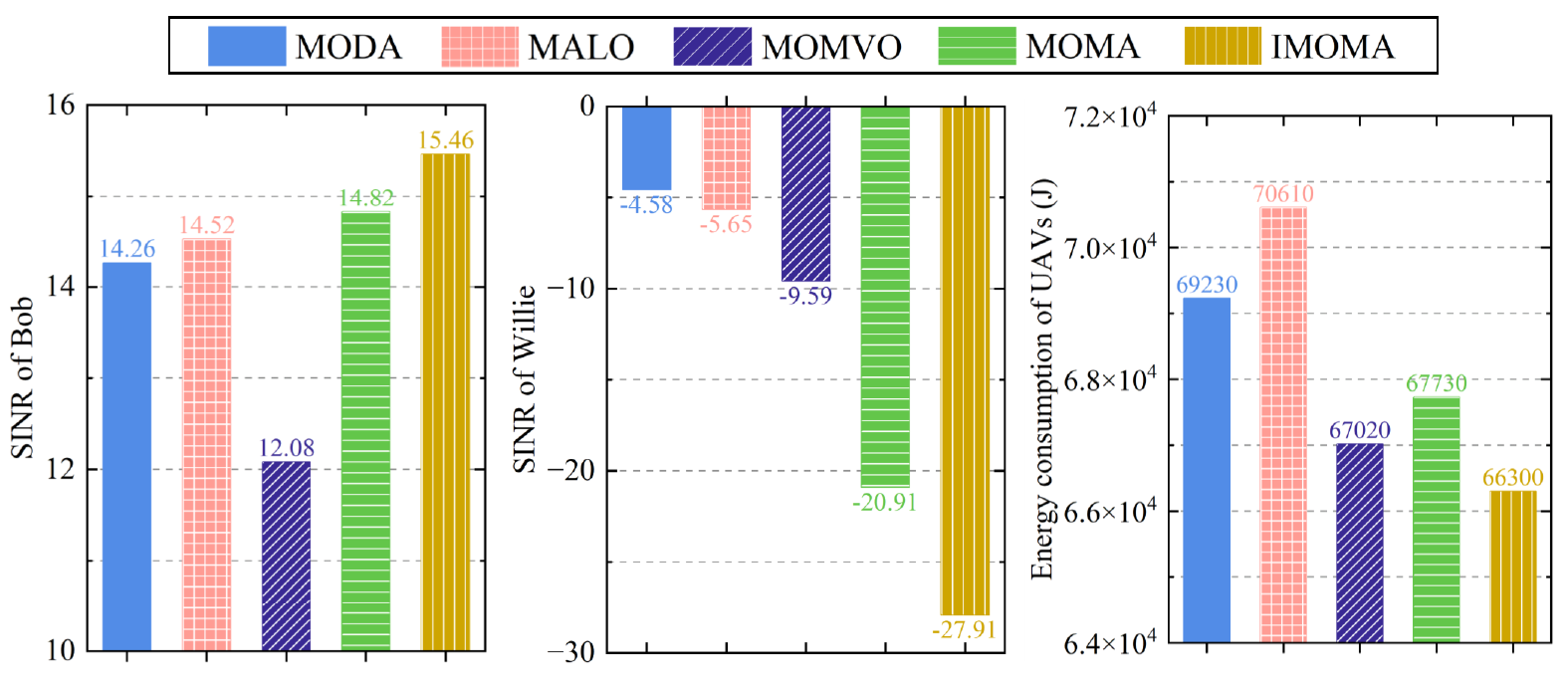}}\\
\caption{The optimization objective values obtained by different algorithms in larger and smaller scale networks of the CB-based approach.}
\label{value_large_and_small}
\end{figure}

 Furthermore, Fig. \ref{value_large_and_small} illustrates the optimization objective values, which are obtained by different algorithms in larger and smaller scale networks of the CB-based approach. As can be seen, the IMOMA achieves the optimal results on the first and second optimization objective values, which indicates that IMOMA can implement the effective communications of the legitimate vessel, and ensures security in the communication process. Notably, the enhancement of optimizing the SINR of Willie is more significant. Moreover, it is clear that the proposed IMOMA is optimal in minimizing the total flight energy consumption of UAVs compared to other algorithms. The abovementioned results further show that IMOMA has strong applicability and robustness during the VAA maritime secure communications. The reason may be that we use the chaotic approach to generate the initial values, which increases the probability that the initial solutions are located around the optimal positions. In addition, the considered hybrid solution update strategies utilize heuristic methods to guide the solution update directions, and it adopts different strategies to update the solutions in different dimensions, thus improving the performance of the proposed IMOMA.

\par
\begin{figure}[!t]
\centering
 \subfigure[The optimization objective values of non-CB approach.]{
\includegraphics[width=3.5in]{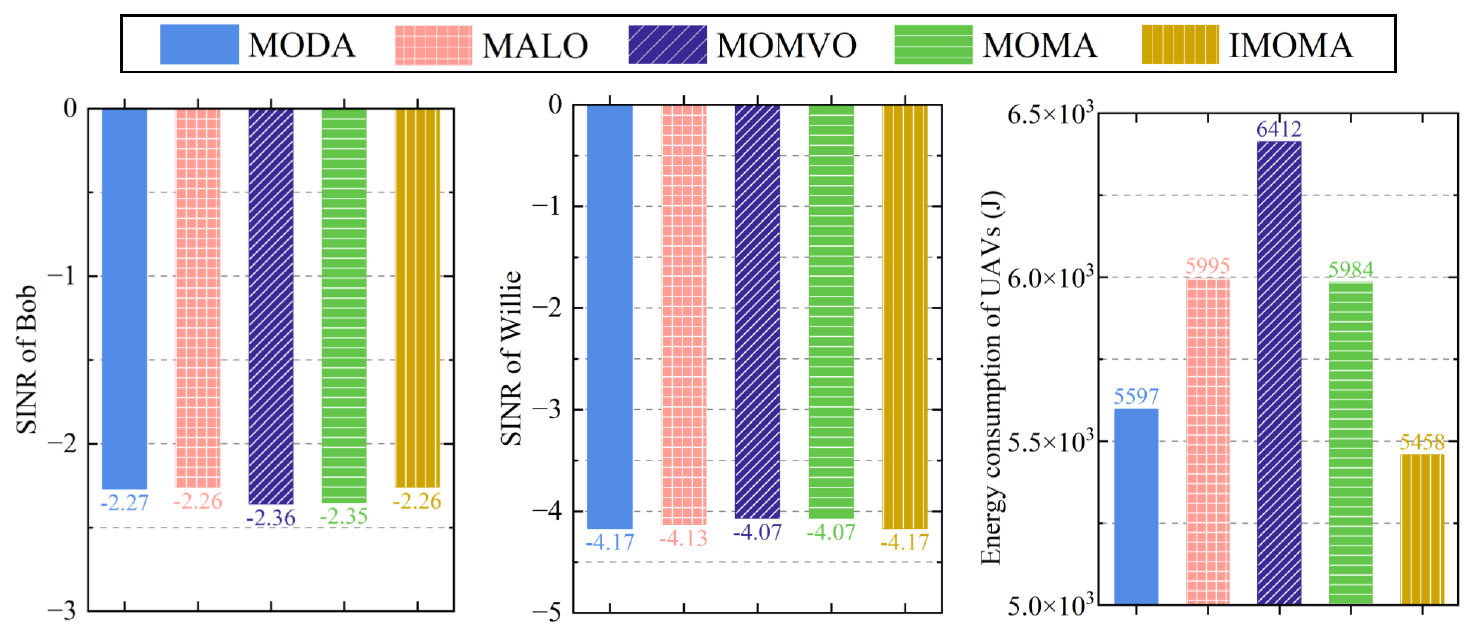}}\\
 \subfigure[The optimization objective values of single CB approach.]{
\includegraphics[width=3.5in]{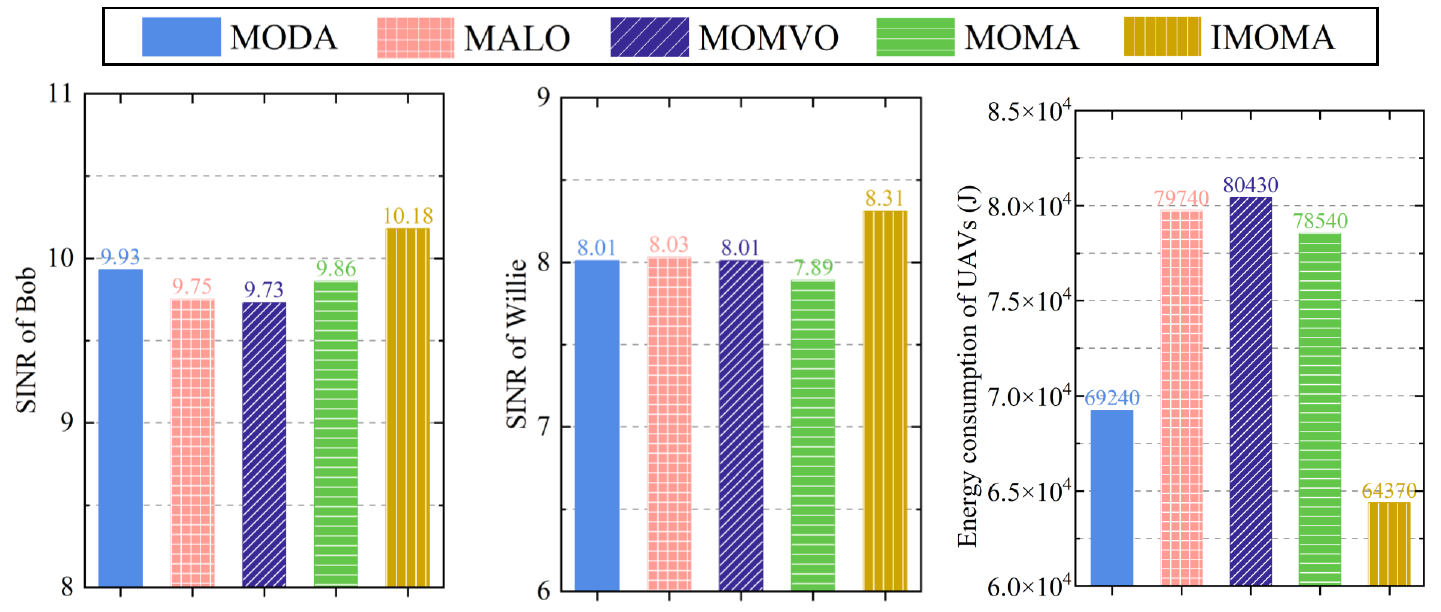}}\\
\caption{The optimization objective values obtained by different algorithms of non-CB and single CB approaches.}
\label{value_algo_compare12}
\end{figure}

 In addition, Fig. \ref{value_algo_compare12} presents the optimization objective values obtained by different algorithms of non-CB and single CB approaches. In Fig. \ref{value_algo_compare12}(a), IMOMA performs well in the three optimization objectives compared to other comparison algorithms, which further confirms the efficiency of IMOMA. In Fig. \ref{value_algo_compare12}(b), IMOMA displays remarkable improvements in the SINR of Bob and energy consumption of UAVs, whereas its greater sensitivity to CB may cause Willie to receive more data signals, resulting in poor results about the SINR of Willie. In general, IMOMA demonstrates strong performance, achieving efficient and secure maritime communications.

\subsubsection{Convergence of the IMOMA}

\par Note that proving convergence is challenging due to the stochastic nature of the algorithm. Moreover, it is difficult to give a direct convergence curve for multi-objective optimization algorithms~\cite{Deb2014}. Thus, we use the \textit{solution distributions}, \textit{inverted generational distance (IGD)}, and \textit{alternative average convergence rate (ACR)} methods to assess the convergence of the IMOMA as follows.

\begin{figure*}
	\centering
	\subfigure[Solution distribution with different iterations.]{\includegraphics[scale=0.065]{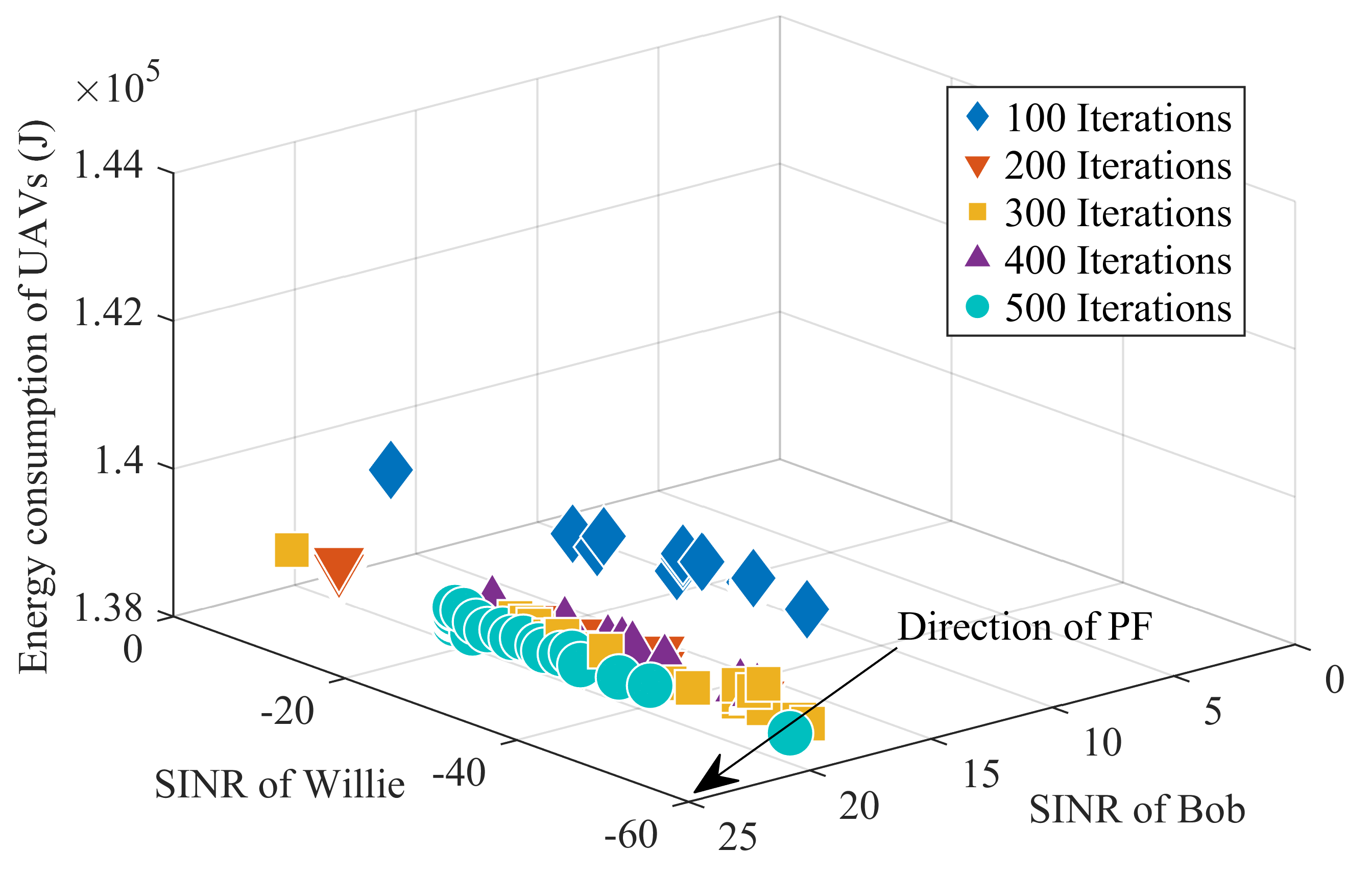}}\hfill
    \subfigure[IGD curve.]{\includegraphics[scale=0.4]{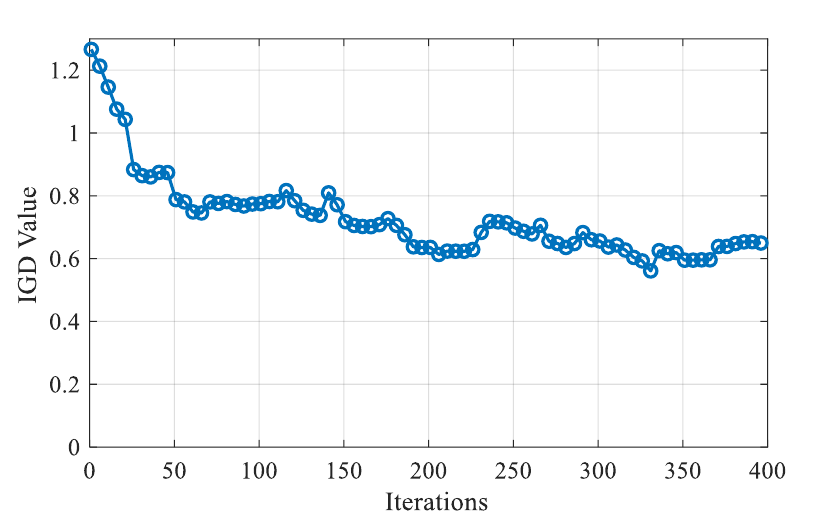}}\hfill
	\subfigure[ACRs of optimization objectives.]{\includegraphics[scale=0.29]{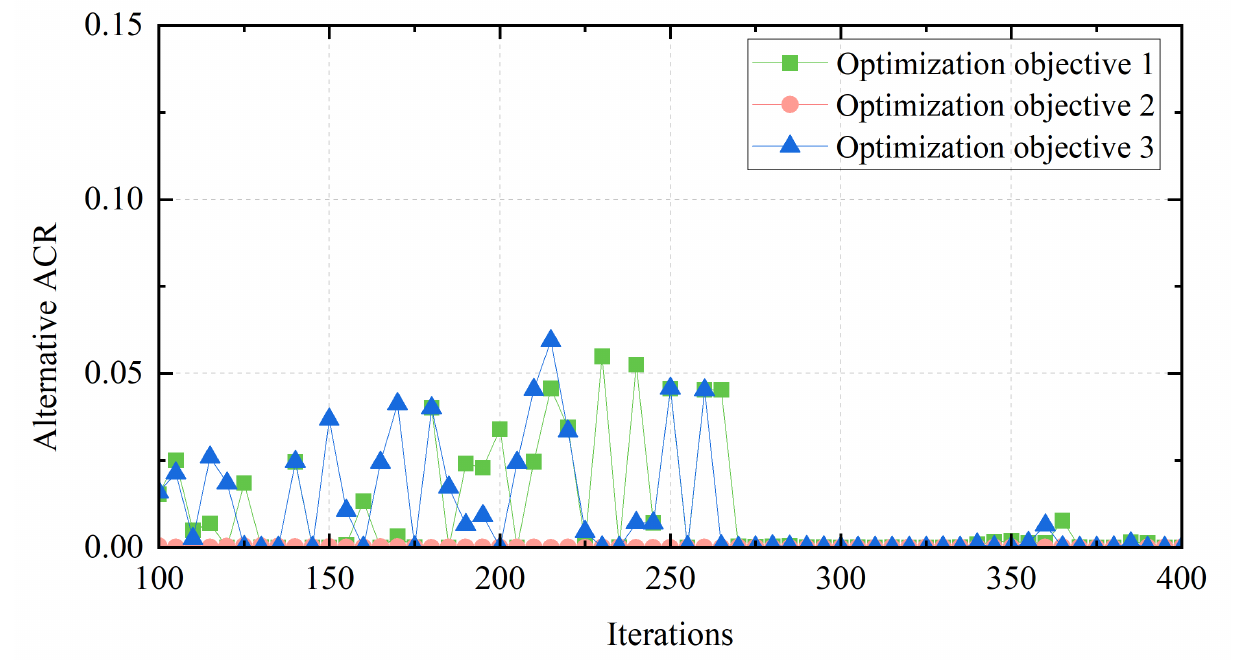}}
	\caption{Convergence analysis of the IMOMA.}
 \label{convergence}
\end{figure*}

\textit{(i) Solution Distributions Method:} We analyze the solution distributions with different iterations, as shown in Fig. \ref{convergence}(a). Specifically, as the number of iterations increases, the solutions gradually approach the Pareto front. When the iterations reach around 300, the distributions begin to overlap, indicating the stabilization of the solution, thereby suggesting convergence of the proposed algorithm.

\textit{(ii) IGD Method:} The IGD measures the average distance between the obtained solution set and the true Pareto front. Since the true Pareto front is often unattainable, we use the non-dominated solutions from multiple experiments to form an approximate Pareto front. Fig. \ref{convergence}(b) presents the IGD curve obtained by IMOMA. As can be seen, the IGD value decreases over iterations, indicating that the solutions are aligning with the Pareto front. After 200 iterations, the IGD stabilizes, indicating that the IMOMA has converged effectively.

\textit{(iii) ACR Method:} We utilize the ACR, which has been shown to effectively reflect convergence performance~\cite{Chen2021a}~\cite{He2015}. For each of the three optimization objectives in IMOMA, we calculate the ACR at each iteration, selecting the best objective value within the Pareto set. Trend plots of the ACR for the three objectives are shown in Fig. \ref{convergence}(c), where we observe that the ACR converges towards 0, indicating convergence.

\par In summary, the results from the solution distribution, IGD, and ACR methods confirm that the proposed IMOMA has effectively converged.




\section{Discussion}
\label{sec:discussion}

\par In this section, the synchronized transmission process of UAVs in the same VAA is further discussed.

\par In each round of G2A communications, with $S$ sensing nodes at the LBS and $N$ UAVs, if each sensing node broadcasts its data individually, $S$ transmissions and $S \times N$ packet receptions are required. To improve energy efficiency, a master node can be selected to collect data from the sensing nodes and multicast the summarized packets to all UAVs. This aggregation reduces ($S \times N$) receptions to ($S + N$) receptions and one data transmission, with an additional $N$ control packets used for the master node selection. The data-sharing process in each round involves the following steps~\cite{Feng2010}:
\begin{itemize}
    \item \textit{Master Node Selection:} UAVs multicast their IDs and residual energy levels to select the UAV with the highest energy as the master node, minimizing communication and ensuring adequate energy for data aggregation and forwarding.
   \item \textit{Master Node ID Sharing:} The selected master node broadcasts its ID to the sensing nodes, so they know where to send their data. This single control packet saves energy compared to multiple transmissions.\item \textit{Data Collection by Master Node:} Sensing nodes transmit their data to the master node, which aggregates the information for efficient distribution.
    \item \textit{Data Multicast to UAVs:} The master node then multicasts the aggregated data to all UAVs, enabling them to perform synchronized beamforming for the final transmission.
\end{itemize}

\par To further reduce communication overhead, master node selections do not need to occur every round. UAVs can exchange energy status information and estimate the number of rounds a master node can sustain, reducing the frequency of selections and enhancing energy efficiency.

\par Furthermore, the overhead of this process is relatively low, approximately 10-20 seconds, which is minimal compared to the time savings in communication and motion achieved by the CB method. Compared to multi-hop approaches, our method saves 50$\%$ to 90$\%$ of the time, with minimal energy consumption, as confirmed by reference~\cite{Feng2010}. This combination of time and energy efficiency makes the method highly applicable to real-world scenarios, where quick decisions and sustainable energy use are essential.

%

\section{Conclusion} 
\label{sec:conclusion}

\par In this paper, the dual UAV cluster-assisted maritime physical layer secure communications via CB were investigated. Specifically, we considered the CB-based dual UAV cluster-assisted maritime secure communication system, which involves maritime long-distance communications and takes into account the security. In the system, one UAV cluster formed an MUVAA relay to forward data signals to the legitimate vessel, and the other UAV cluster formed an MUVAA jammer to send jamming signals to the eavesdropper. Moreover, taking into account the conflicting objectives, we formulated the SEMCMOP. Then, to resolve the complex NP-hard and large-scale problem, we proposed the IMOMA with chaotic solution initialization and hybrid solution update strategies. Simulation results showed that the CB-based method is significantly better than that of the non-CB, single CB, and multi-hop approaches, which means that CB is suitable for long-distance maritime secure communications. Moreover, comparison results indicated the proposed IMOMA outperforms several comparison algorithms and is more suitable for CB-based maritime long-distance secure communication scenarios. Future work can extend the results of this study by incorporating real-time variations in vessel positions and adopting more adaptive DRL algorithms, enhancing the ability of system dynamics to autonomously adjust to real-world scenarios. Additionally, exploring high-altitude communication platforms, such as HAPs, could further enhance system diversity.

\ifCLASSOPTIONcaptionsoff
  \newpage
\fi

%


\begin{IEEEbiography}
[{\includegraphics[width=1in,height=1.25in,clip,keepaspectratio]{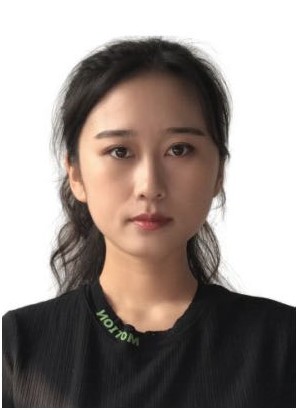}}]
{Jiawei Huang} received a BS degree in Software Engineering from Dalian Jiaotong University, and a MS degree in Software Engineering from Jilin University in 2019 and 2024, respectively. She is currently studying Computer Science at Jilin University to get a Ph.D. degree. Her current research interests are UAV networks and optimization.
\end{IEEEbiography}

\begin{IEEEbiography}
[{\includegraphics[width=1in,height=1.25in,clip,keepaspectratio]{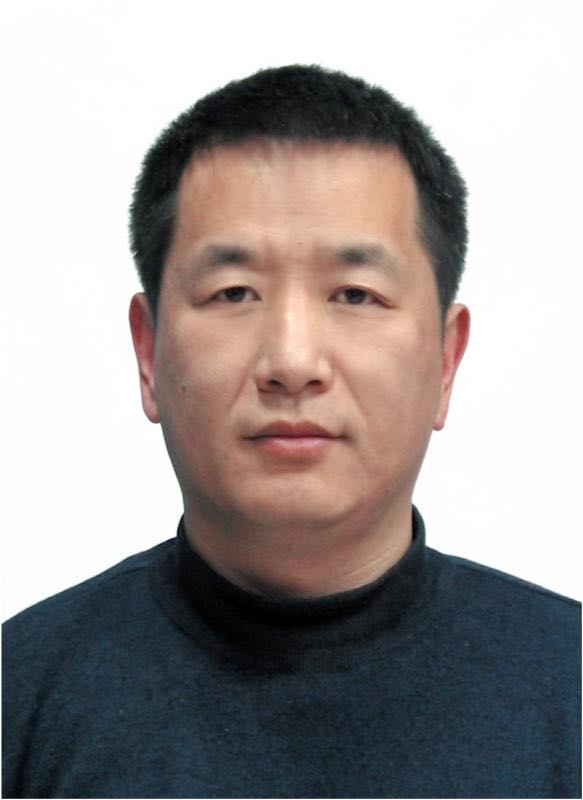}}]
{Aimin Wang} received Ph.D. degree in Communication and Information System from Jilin University. He is currently a professor at Jilin University. His research interests are wireless sensor networks and QoS for multimedia transmission.
\end{IEEEbiography}

\begin{IEEEbiography}
[{\includegraphics[width=1in,height=1.25in,clip,keepaspectratio]{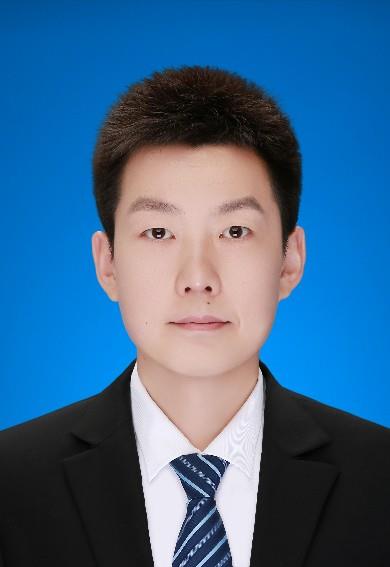}}]
{Geng Sun} (Senior Member, IEEE) received the B.S. degree in communication engineering from Dalian Polytechnic University, in 2007, and the Ph.D. degree in computer science and technology from Jilin University, in 2018. He was a Visiting Researcher with the School of Electrical and Computer Engineering, Georgia Institute of Technology, USA. He is a Professor in College of Computer Science and Technology at Jilin University, and his research interests include wireless networks, UAV communications, collaborative beamforming and optimizations. 
\end{IEEEbiography}

\begin{IEEEbiography}
[{\includegraphics[width=1in,height=1.25in,clip,keepaspectratio]{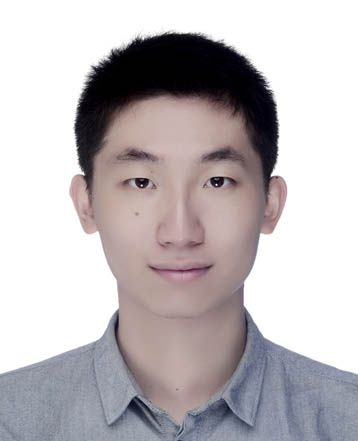}}]
{Jiahui Li} received the B.S. degree in Software Engineering, and the Ph.D. degree in Computer Science and Technology from Jilin University, Changchun, China, in 2018 and 2024, respectively. He is currently a postdoctoral researcher in College of Computer Science and Technology at Jilin University. His current research focuses on UAV networks, antenna arrays, and optimization.
\end{IEEEbiography}

\begin{IEEEbiography}
[{\includegraphics[width=1in,height=1.25in,clip,keepaspectratio]{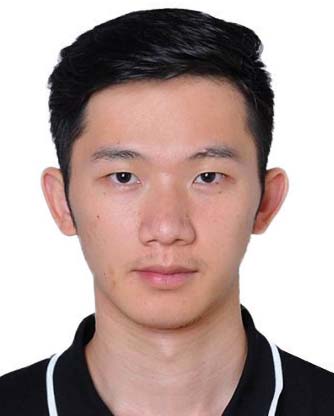}}]{Jiacheng Wang} received the Ph.D. degree from the School of Communication and Information Engineering, Chongqing University of Posts and Telecommunications, Chongqing, China. He is currently a Research Associate in computer science and engineering with Nanyang Technological University, Singapore. His research interests include wireless sensing, semantic communications, and metaverse.
\end{IEEEbiography}

\begin{IEEEbiography}
[{\includegraphics[width=1in,height=1.25in,clip,keepaspectratio]{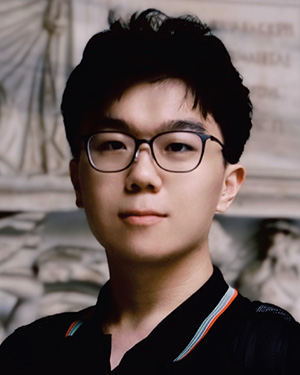}}]{Hongyang Du} received the B.Sc. degree from Beijing Jiaotong University, Beijing, China, in 2021, and the Ph.D. degree from Nanyang Technological University, Singapore, in 2024. He is currently an assistant professor at the Department of Electrical and Electronic Engineering, the University of Hong Kong (HKU), and the Principal Investigator (PI) of the Network Intelligence and Computing Ecosystem (NICE) Lab. His research interests include semantic communications, generative AI, and resource allocation.
\end{IEEEbiography}

\begin{IEEEbiography}
[{\includegraphics[width=1in,height=1.25in,clip,keepaspectratio]{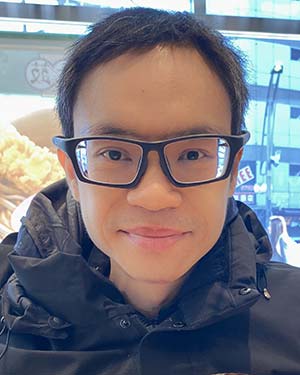}}]{Dusit Niyato} (Fellow, IEEE) received the BEng degree from the King Mongkuts Institute of Technology Ladkrabang, Thailand, in 1999, and the PhD degree in electrical and computer engineering from the University of Manitoba, Canada, in 2008. He is currently a professor with the School of Computer Science and Engineering, Nanyang Technological University, Singapore. His research interests include the Internet of Things (IoT), machine learning, and incentive mechanism design.
\end{IEEEbiography}

\end{document}